\documentclass[useAMS,usenatbib]{mn2e}

\usepackage{natbib, aas_macros}
\usepackage{graphicx}

\newcommand {\apgt} {\ {\raise-.5ex\hbox{$\buildrel>\over\sim$}}\ }
\newcommand {\aplt} {\ {\raise-.5ex\hbox{$\buildrel<\over\sim$}}\ } 

\title[The moment of core collapse in star clusters]{The moment of core collapse in star clusters  with a mass function}

\author[M. S. Fujii and S. Portegies Zwart]
{M. S. Fujii$^{1,2}$
\thanks{E-mail: michiko.fujii@nao.ac.jp (MSF); spz@strw.leidenuniv.nl (SPZ)} 
and S. Portegies Zwart$^{2}$\footnotemark[1]\\
$^{1}$Division of Theoretical Astronomy, National Astronomical Observatory of Japan, 2-21-1 Osawa, Mitaka, Tokyo, 181-8588, Japan\\
$^{2}$Leiden Observatory, Leiden University, NL-2300RA Leiden, The Netherlands}
\begin{document}

\date{Accepted 1988 December 15. Received 1988 December 14; in original form 1988 October 11}

\pagerange{\pageref{firstpage}--\pageref{lastpage}} \pubyear{2002}

\maketitle

\label{firstpage}

\begin{abstract}
Star clusters with multi-mass components dynamically evolve faster
than those modeled with equal-mass components.  Using a series of
direct $N$-body simulations, we investigate the dynamical evolution of
star clusters with mass functions, especially their core collapse
time. Multi-mass clusters tend to behave like systems with a smaller
number of particles, which we call the effective number of particles
($N_{\rm eff}$) and for which $N_{\rm eff} = M/m_{\rm max}$ (here $M$
and $m_{\rm max}$ are the total cluster mass and the mass of the most
massive star in the cluster, respectively).  We find that the time of
core collapse is inversely proportional to the mass of the most
massive star in the cluster and analytically confirm that this is
because the core collapse of clusters with a mass function proceeds on
the dynamical friction timescale of the most massive stars.  As the
mass of the most massive star increases, however, the core-collapse
time, which is observed as a core bounce of the cluster core from the
evolution of the core density or core radius, becomes ambiguous.  We
find that in that case the total binding energy of the hard binaries
gives a good diagnosis for determining the moment of
core collapses.  Based on the results of our simulations, we
argue that the core bounce becomes ambiguous when the mass of the most
massive star exceeds 0.1\% of the total mass of the cluster.
\end{abstract}


\section{Introduction}

Star clusters are collisional systems with a negative heat capacity,
and therefore they dynamically evolve to eventually reach core
collapse.  The process to core collapse is simply described using a
semi-analytic treatment of the energy transfer from the cluster core
to the outer part of the cluster \citep{1968MNRAS.138..495L}, and it
was confirmed using various methods such as gaseous models
\citep{1970MNRAS.150...93L}, Fokker-Planck simulations 
\citep{1979ApJ...234.1036C,1980ApJ...242..765C,1980MNRAS.191..483L,
1980PASJ...32..213I}, 
and direct $N$-body simulations \citep{1974A&A....37..183A}.  
Because of the high
stellar density of the core during its collapse, dynamical binaries
form in this phase \citep{1971ApJ...164..399S,1974A&A....35..237A}.
Once binaries form in the core, they generate energy which is
transported outward by interactions with other cluster members
\citep{1975MNRAS.173..729H,1983ApJ...272L..29H}. Due to the binary
heating the core finally bounces, and then the core oscillates when
the number of stars is sufficiently large \citep{1987ApJ...313..576G}.
Such gravothermal oscillation was first found using gaseous models
\citep{1983MNRAS.204P..19S,1984MNRAS.208..493B}, but the behavior was
later confirmed also using Fokker-Planck
\citep{1989ApJ...342..814C} and direct $N$-body
\citep{1996ApJ...471..796M} simulations.

The time between cluster birth and the moment of core collapse (what
we will call the core-collapse time, or $t_{\rm cc}$) is proportional
to the two-body relaxation time at the half-mass radius, $t_{\rm rh}$
\citep{1987degc.book.....S}. For clusters in which all stars have the
same mass, the core collapses at $t_{\rm cc}\simeq 15$--$20t_{\rm rh}$. This
result has a theoretical background and was confirmed with simulations
using a wide variety of techniques: Monte-Carlo methods
\citep{1975ApJ...200..339S}, Fokker-Planck calculations
\citep{1980ApJ...242..765C,1995PASJ...47..561T}, and direct $N$-body 
simulations \citep{1996ApJ...471..796M}. With a spectrum of stellar masses,
however, the core-collapse time becomes much shorter than when all
stars have the same mass \citep{1984PASJ...36..391I}.  From direct
$N$-body simulations, it is obtained that $t_{\rm cc}\simeq 0.2t_{\rm
  rh}$ with a realistic mass function \citep{2002ApJ...576..899P}.

\citet{2004ApJ...604..632G} performed a large number of Monte Carlo
simulations covering a wide range of the concentration parameters and
mass functions.  They showed that the core-collapse time scales with
the central relaxation time, $t_{\rm rc}$, rather than with $t_{\rm
  rh}$.  Here the central relaxation time is defined as
\begin{eqnarray}
t_{\rm rc} \equiv
\frac{0.065 \sigma_{\rm c, 3D}^3}{G^2 \langle m\rangle \rho_{\rm c} \ln \Lambda}.
\label{eq:tcrlx}
\end{eqnarray}
Here $\sigma _{\rm c, 3D}$ and $\rho_{\rm c}$ are the central
three-dimensional velocity dispersion and core density, respectively
\citep{1987degc.book.....S,2004ApJ...604..632G}.  The Coulomb
logarithm, $\ln \Lambda$, is written as $\ln \gamma N$, where $N$ is
the total number of stars. For the central relaxation time of star
clusters, it is numerically obtained that $\gamma = 0.015$
\citep{1996MNRAS.279.1037G} (the classic theoretical value for $t_{\rm
  rh}$ with single-mass components is $\gamma = 0.11$
\cite{1987degc.book.....S}).

In the case with a power-law mass function in which $m_{\rm max}$ and
$\langle m \rangle$ are the maximum mass and the mean mass of the
stellar mass distribution, \citet{2004ApJ...604..632G} found that the
core-collapse time $t_{\rm cc}\propto (m_{\rm max}/\langle m
\rangle)^{-1.3}$. They also demonstrated the existence of a minimum to
the core-collapse time, which is $t_{\rm cc}/t_{\rm rc} = 0.15$. This
value is also seen in \citet{2012ApJ...752...43G}, in which they use
the same Monte Carlo code that was adopted by
\citet{2004ApJ...604..632G}, but with a wider range of initial
conditions including initially mass-segregated models.  The arguments
for the particular exponent (being $-1.3$) and the minimum to the
core-collapse time, however, was not discussed and remains unclear.

In this paper we show the results of core collapse simulations of star
clusters with power-law mass functions using direct $N$-body
simulations.  We find that the core collapse time scales $t_{\rm
  cc}\propto (m_{\rm max}/\langle m \rangle)^{-1}$, contrary to the
earlier finding of \citet{2004ApJ...604..632G}, but we support our
finding with analytic arguments.  We further argue that core collapse
is driven by the sinking of the most massive stars to the cluster
center, by dynamical friction (as was suggested in
\cite{2002ApJ...576..899P}).  The time to the core-collapse 
then corresponds to the time required for the most massive star to
reach the cluster center.  For the most extreme mass functions, the
core-collapse time then naturally depends on the crossing time
of the system rather than the dynamical friction timescale. 

The core bounce becomes less pronounced for larger
values of $m_{\rm max}/\langle m \rangle$, and this is qualitatively
understood from the dynamical evolution being driven by the most
massive stars in the cluster. For a mass ratio $M/m_{\rm max} \aplt
10^3$, core collapse becomes hard to determine, and it even becomes
indistinguishable for $M/m_{\rm max} \aplt 100$, because in those
cases the core will eventually be composed of only a few massive
stars, almost irrespective of the total number of stars in the
cluster. In such a case, however, the binding energy of the 
hardest binary gives us a good indication to detect the moment
of core collapse.

\section{$N$-body simulations and the initial conditions}

We performed a series of $N$-body simulations using King models
\citep{1966AJ.....71...64K} with a non-dimensional concentration
parameter, $W_0$, of 3 and 6 as initial density profiles.  Hereafter,
we adopt $N$-body units in which, $G=M=-4E=1$, where $G$, $M$, and $E$
are the gravitational constant, the total mass, and the total energy
of the cluster, respectively \citep{1986LNP...267..233H}\footnote{See
  also {\tt http://en.wikipedia.org/wiki/N-body\_units}.}. We 
construct the initial particle distributions using NEMO 
\citep{1995ASPC...77..398T}.
In table \ref{tb:model_cl}
we summarize the initial conditions for the runs.  For each simulation
we adopted $N$ particles from $N=2048$ (2k), 8192 (8k), 32768 (32k) to
$N=1310172$ (128k), with a power-law mass function with exponent
$-\alpha$ and an upper-mass limit of $m_{\rm max}$.  The value of
$m_{\rm max}/\langle m \rangle \equiv f_{\rm max}$ is varied from 1.0
(equal mass) to 517, but for models with large-$N$ models we adopted a
large value of $m_{\rm max}$ because of the calculation time.  For the
mass function exponent $\alpha$, we adopted $\alpha=2.35$
\citep{1955ApJ...121..161S}, 1.7, and 1.2. In
Tables~\ref{tb:model_cl_w3} and \ref{tb:model_cl_w6} we summarize the
simulation results.

\begin{table}
\begin{center}
\caption{Properties of King Models\label{tb:model_cl}}
\begin{tabular}{cccccc}\hline\hline
$W_{0}$ & $\rho_{\rm c}$ & $\sigma_{\rm c, 1D}$ & $r_{\rm c}$ & $r_{\rm h}$ & $M_{\rm c}$    \\ \hline
3  & 0.652 & 0.518 & 0.543  & 0.839 & 0.238  \\
6  & 2.11 & 0.503  & 0.293  & 0.804 & 0.117  \\
\hline
\end{tabular}
\end{center}

 \medskip
Definitions of quantities listed in this table: 
$W_{0}$ is a non-dimensional concentration
parameter for King models; $\rho_{\rm c}$ is the core density; $\sigma_{\rm c, 1D}$ is the velocity dispersion in the core; $r_{\rm c}$ is the core radius; $r_{\rm h}$is the half-mass radius; $M_{\rm c}$  is the mass within the core. 
\end{table}

All simulations are performed using the sixth-order
predictor-corrector Hermite scheme \citep{2008NewA...13..498N} running
on GPU using the Sapporo2 library
\citep{2008NewA...13..103B,2012EPJST.210..201B} and also on CPU
clusters using the two-dimensional parallelization scheme by
\citet{2006astro.ph..6105N}.    We used a
time step criterion \citep{2008NewA...13..498N} with accuracy
parameter, $\eta=$0.1--0.3.  The energy error is $\aplt 10^{-4}$ for
equal-mass models and $\aplt 10^{-5}$ for all simulations over the
entire duration of the simulation.  
For the models with $f_{\rm max} \aplt 2$, the energy error tends
to become larger compared to the models with  $f_{\rm max} > 2$,
especially after the formation of a binary of $\sim 10 kT$.
(We express binding energies in terms
of $kT \equiv \langle m \rangle \sigma_{\rm 1D} ^2$, where
$\sigma_{\rm 1D}$ is the one-dimensional velocity dispersion of the
cluster, $3NkT/2$ is the initial kinetic energy of the entire system.)
If we try to adopt small timesteps in order to maintain less 
energy error, the time step of the calculations dropped below 
$\sim10^{-13}$. To prevent such small timesteps, which have unpleasant 
consequences
for the performance, we also performed simulations adopting a small
softening $\epsilon$; for the simulations with $W_0=3$ we adopted
$\epsilon = 1/(200N)$ and $\epsilon = 1/(130N)$ for models with
$W_0=6$. With this softening we are able to resolve binaries with a
semi-major axis of a $1200kT$ and $780kT$ for the simulations with
$W_0=3$ and $W_0=6$, respectively. With softening length, energy 
error is $\aplt 10^{-5}$ throughout the simulations.

\begin{table*}
\begin{center}
\caption{Models with $W_{0}=3$\label{tb:model_cl_w3}}
\begin{tabular}{lcccccccc}\hline\hline
Model & $W_{0}$ & $\alpha$ & $m_{\rm min}/\langle m \rangle$ & $m_{\rm max}/\langle m \rangle$ & $N$ & $\epsilon$ & $N_{\rm run}$   \\ \hline
w3-2k-eq       & 3 & - & 1.0 & 1.0 & 2k  & 0 & 5  \\
w3-2k-eq-soft  & 3 & - & 1.0 & 1.0 & 2k & $1/(200N)$  & 7\\
w3-8k-eq-soft  & 3 & - & 1.0 & 1.0 & 8k  & $1/(200N)$  & 1 \\
\\
w3-2k-m2-Sal   & 3 & 2.35 & 0.607 & 2.02 & 2k   & 0  & 7 \\
w3-2k-m2-Sal-soft  & 3 & 2.35 & 0.607 & 2.02 & 2k   & $1/(200N)$ & 8 \\
w3-8k-m2-Sal-soft  & 3 & 2.35 & 0.607& 2.02 & 8k   & $1/(200N)$  & 1 \\
\\
w3-2k-m8-Sal  & 3 & 2.35 & 0.391 & 8.07 & 2k   & 0  & 10 \\
w3-8k-m8-Sal  & 3 & 2.35 & 0.391 & 8.07 & 8k    & 0 & 1  \\
w3-32k-m8-Sal  & 3 & 2.35 & 0.391 & 8.07 & 32k    & 0 & 3 \\
\\
w3-2k-m32-Sal  & 3 & 2.35 & 0.329 & 32.3 & 2k   & 0  & 10 \\
w3-8k-m32-Sal  & 3 & 2.35 & 0.329 & 32.3 & 8k   & 0  & 1 \\
w3-32k-m32-Sal  & 3 & 2.35 & 0.329 & 32.3 & 32k  & 0   & 8 \\
\\
w3-2k-m129-Sal  & 3 & 2.35 & 0.296 & 129.2 & 2k   & 0  & 10 \\
w3-8k-m129-Sal  & 3 & 2.35 & 0.296 & 129.2 & 8k   & 0  & 1 \\
w3-32k-m129-Sal  & 3 & 2.35 & 0.296 & 129.2 & 32k   & 0  & 8 \\
w3-128k-m129-Sal  & 3 & 2.35 & 0.279 & 516.6 & 128k  & 0  & 7 \\
\\
w3-2k-m258-Sal  & 3 & 2.35 & 0.286 & 258.3 & 2k  & 0   & 2 \\
w3-8k-m258-Sal  & 3 & 2.35 & 0.286  & 258.3 & 8k   & 0  & 1 \\
w3-32k-m258-Sal  & 3 & 2.35 & 0.286 & 258.3 & 32k  & 0  & 1 \\
\\
w3-2k-m517-Sal  & 3 & 2.35 & 0.279 & 516.6 & 2k  & 0 & 7\\
w3-8k-m517-Sal  & 3 & 2.35 & 0.279  & 516.6 & 8k  & 0  & 1 \\
w3-32k-m517-Sal  & 3 & 2.35 & 0.279 & 516.6 & 32k  & 0  & 8 \\
w3-128k-m517-Sal & 3 & 2.35 & 0.279 & 516.6 & 128k  & 0  & 8 \\
\\
w3-2k-m2-a1.2-soft  & 3 & 1.2 & 0.442 & 2.02 & 2k & $1/(200N)$ &  1 \\
w3-8k-m2-a1.2-soft  & 3 & 1.2 & 0.442 & 2.02 & 8k  & $1/(200N)$  & 1 \\
\\
w3-2k-m8-a1.2  & 3 & 1.2 & 0.0336 & 8.07 & 2k   & 0   & 1 \\
w3-8k-m8-a1.2  & 3 & 1.2 & 0.0336 & 8.07 & 8k   & 0   & 1 \\
w3-32k-m8-a1.2  & 3 & 1.2 & 0.0336 & 8.07 & 32k   & 0  & 4 \\
\\
w3-2k-m32-a1.2  & 3 & 1.2 & $5.28\times 10^{-4}$ & 32.3 & 2k & 0  & 1  \\
w3-8k-m32-a1.2  & 3 & 1.2 & $5.28\times 10^{-4}$ & 32.3 & 8k & 0  & 1 \\
w3-32k-m32-a1.2  & 3 & 1.2 & $5.28\times 10^{-4}$ & 32.3 & 32k & 0 & 4 \\
\\
w3-2k-m129-a1.2  & 3 & 1.2 & $5.16\times 10^{-6}$ & 129.2 & 2k & 0  & 1 \\
w3-8k-m129-a1.2  & 3 & 1.2 & $5.16\times 10^{-6}$ & 129.2 & 8k & 0  & 1  \\
w3-32k-m129-a1.2  & 3 & 1.2 & $5.16\times 10^{-6}$ & 129.2 & 32k & 0 & 4 \\
\\
w3-2k-m258-a1.2  & 3 & 1.2 & $2.13\times 10^{-8}$ & 258.3 & 2k & 0  & 1 \\
w3-8k-m258-a1.2  & 3 & 1.2 & $2.13\times 10^{-8}$ & 258.3 & 8k & 0  & 1 \\
\\
w3-8k-m517-a1.2  & 3 & 1.2 & $1.23\times 10^{-8}$ & 516.6 & 8k & 0  & 1 \\
w3-32k-m517-a1.2  & 3 & 1.2 & $1.23\times 10^{-8}$ & 516.6 & 32k & 0  & 4 \\
\\
w3-2k-m2-a1.7-soft & 3 & 1.7 & 0.525 & 2.02 & 2k  & $1/(200N)$ & 1 \\
w3-2k-m8-a1.7  & 3 & 1.7 & 0.192 & 8.07 & 2k  & 0  & 1  \\
w3-2k-m32-a1.7  & 3 & 1.7 & 0.0856  & 32.3 & 2k  & 0  & 1  \\
w3-8k-m32-a1.7  & 3 & 1.7 & 0.0856  & 32.3 & 8k  & 0  & 1 \\
w3-2k-m129-a1.7  & 3 & 1.7 & 0.0403  & 129.2 & 2k  & 0  & 1 \\
w3-8k-m258-a1.7  & 3 & 1.7 & 0.0304  & 258.3 & 8k  & 0  & 1 \\
\hline
\end{tabular}
\end{center}

 \medskip
$N_{\rm run}$ is the number of runs with the same initial parameters, but 
realized with different random seeds.
\end{table*}

\begin{table*}
\begin{center}
\caption{Models with $W_{0}=6$ \label{tb:model_cl_w6}}
\begin{tabular}{lccccccc}\hline\hline
Model & $W_{0}$ & $\alpha$ & $m_{\rm min}/\langle m \rangle$ & $m_{\rm max}/\langle m \rangle$ & $N$ & $\epsilon$ & $N_{\rm run}$\\ \hline
w6-2k-eq      & 6 & - & 1.0 & 1.0 & 2k  & 0 &  5 \\
w6-2k-eq-soft  & 6 & - & 1.0 & 1.0 & 2k  & $1/(130N)$ &  1 \\
w6-8k-eq-soft  & 6 & - & 1.0 & 1.0 & 8k  & $1/(130N)$ &  1 \\
w6-32k-eq  & 6 & - & 1.0 & 1.0 & 32k  & 0 &  1 \\
\\
w6-2k-m2-Sal-soft  & 6 & 2.35 & 0.607 & 2.02 & 2k & $1/(130N)$  &1 \\
w6-8k-m2-Sal-soft  & 6 & 2.35 & 0.607& 2.02 & 8k  & $1/(130N)$ &1 \\
\\
w6-2k-m8-Sal  & 6 & 2.35 & 0.391 & 8.07 & 2k & 0  &1\\
w6-8k-m8-Sal  & 6 & 2.35 & 0.391 & 8.07 & 8k  & 0  &1\\
w6-32k-m8-Sal  & 6 & 2.35 & 0.391 & 8.07 & 32k  & 0 &4\\
\\
w6-2k-m32-Sal  & 6 & 2.35 & 0.329 & 32.3 & 2k  & 0 &1\\
w6-8k-m32-Sal  & 6 & 2.35 & 0.329 & 32.3 & 8k  & 0  &1\\
w6-32k-m32-Sal  & 6 & 2.35 & 0.329 & 32.3 & 32k  & 0 &4\\
\\
w6-2k-m129-Sal  & 6 & 2.35 & 0.296 & 129.2 & 2k  & 0  &1\\
w6-8k-m129-Sal  & 6 & 2.35 & 0.296 & 129.2 & 8k & 0  &1\\
w6-32k-m129-Sal  & 6 & 2.35 & 0.296 & 129.2 & 32k & 0 &4\\
\\
w6-2k-m258-Sal  & 6 & 2.35 & 0.286 & 258.3& 2k  & 0  &4\\
\\
w6-2k-m517-Sal  & 6 & 2.35 & 0.279 & 516.6 & 2k  & 0  &1\\
w6-8k-m517-Sal  & 6 & 2.35 & 0.279  & 516.6 & 8k  & 0 &1\\
w6-32k-m517-Sal  & 6 & 2.35 & 0.279 & 516.6 & 32k & 0  &4\\
\\
w6-2k-m2-a1.2  & 6 & 1.2 & 0.442 & 2.02 & 2k & 0  &1\\
w6-8k-m2-a1.2  & 6 & 1.2 & 0.442 & 2.02 & 8k  & 0  &1\\
\\
w6-2k-m8-a1.2  & 6 & 1.2 & 0.0336 & 8.07 & 2k  & 0  &1\\
w6-8k-m8-a1.2  & 6 & 1.2 & 0.0336 & 8.07 & 8k  & 0  &1\\
\\
w6-2k-m32-a1.2  & 6 & 1.2 & $5.28\times 10^{-4}$ & 32.3 & 2k  & 0  &1 \\
w6-8k-m32-a1.2  & 6 & 1.2 & $5.28\times 10^{-4}$ & 32.3 & 8k  & 0  &1\\
\\
w6-2k-m129-a1.2  & 6 & 1.2 & $5.16\times 10^{-6}$ & 129.2 & 2k  & 0 &1\\
w6-8k-m129-a1.2  & 6 & 1.2 & $5.16\times 10^{-6}$ & 129.2 & 8k  & 0  &1\\
w6-32k-m129-a1.2  & 6 & 1.2 & $5.16\times 10^{-6}$ & 129.2 & 32k & 0  &1\\
\\
w6-8k-m517-a1.2  & 6 & 1.2 & $1.23\times 10^{-8}$ & 516.6 & 8k  & 0  &1\\
w6-32k-m517-a1.2  & 6 & 1.2 & $1.23\times 10^{-8}$ & 516.6 & 32k  & 0  &1 \\
\hline
\end{tabular}
\end{center}
\end{table*}

\section{Core collapse in multi-mass clusters}

The core-collapse time is usually determined by the moment of core 
bounce, which is seen in the time evolution of the core radius or core
density. For some models, it is difficult to distinguish the core
collapse, because there does not seem to be a peak in the density
evolution or a depression in the core radius. Another indicator for
determining the moment of core collapse is by monitoring the evolution
of the binding energy of dynamically formed binaries. During the core
collapse, hard binaries form in the cluster. They are hardened by
three-body encounters in the cluster core and eventually generate the 
energy for the core bounce.  In this section, we present the evolution 
of the
core radius and the density from the simulations, and then we discuss
the relation between the core evolution and the dynamically formed
binaries in order to provide an objective determination of the moment
of core collapse and define the core collapse.

\subsection{The evolution of core density and radius}

\begin{figure*}
\begin{center}
\includegraphics[width=80mm]{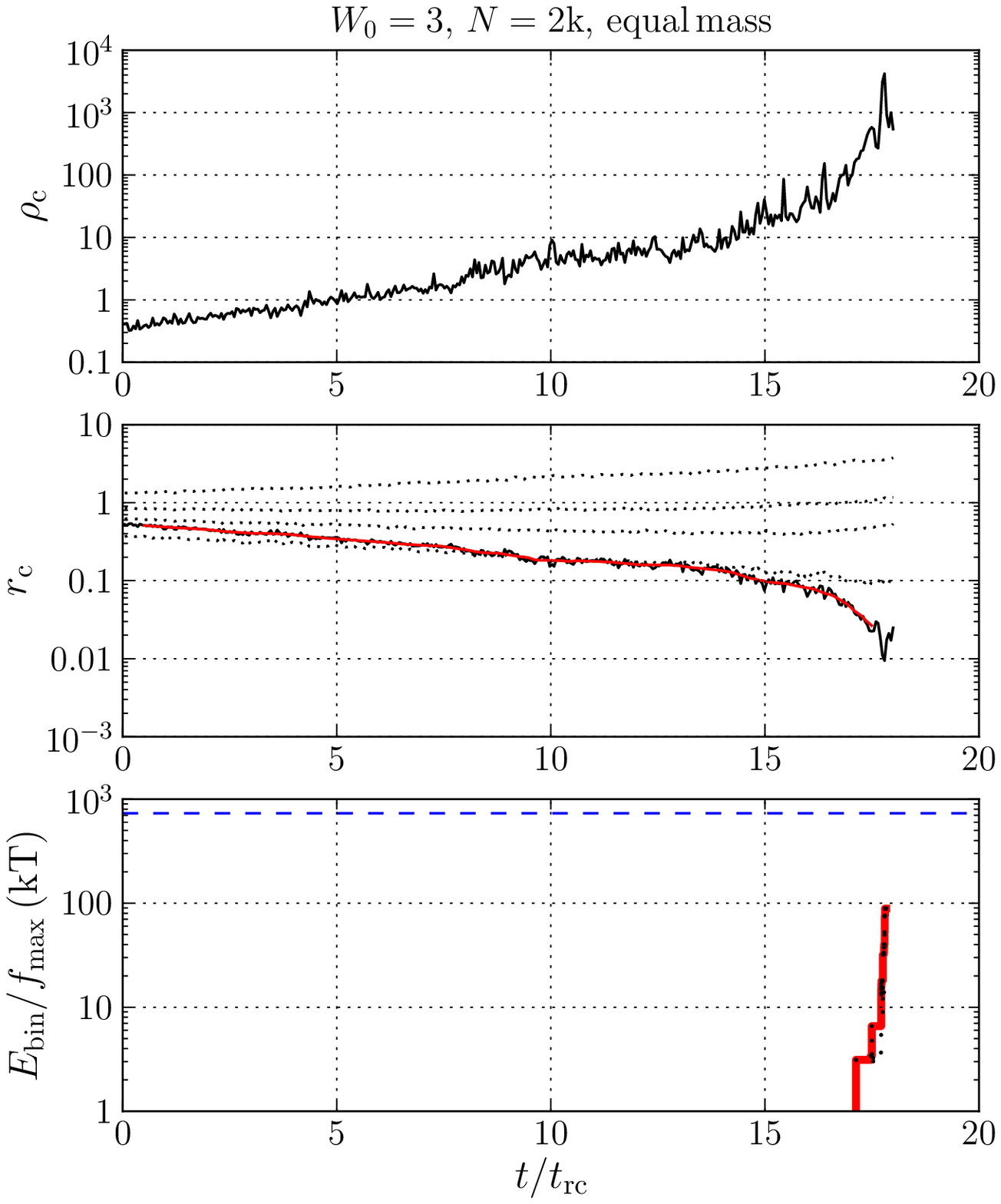}
\includegraphics[width=80mm]{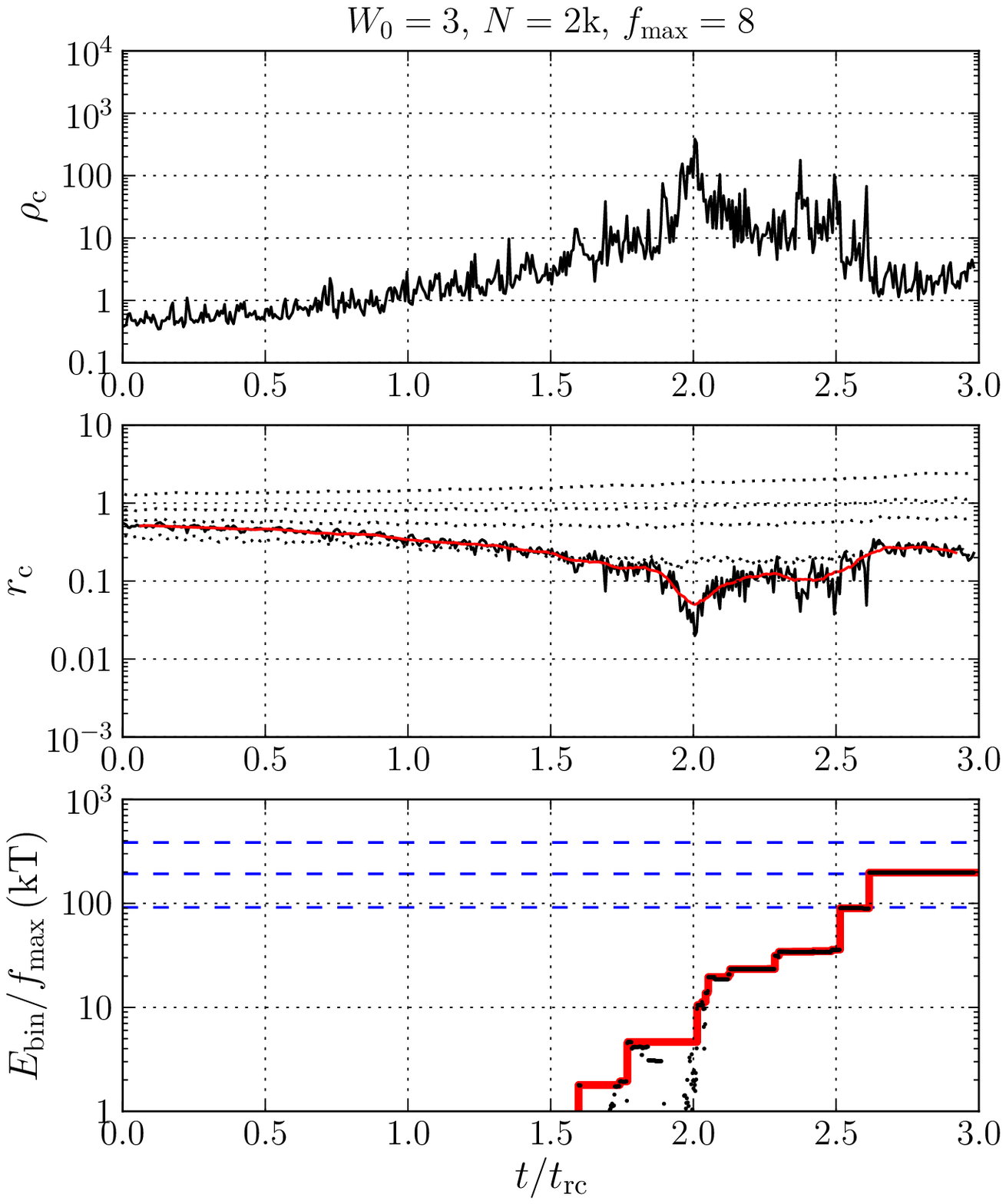}
\end{center}
\caption{Time evolution of the cluster parameters.  Top panel: the
  evolution of the core density.  Middle panel: the evolution of the core
  radius (solid black curve) and the 10\%, 30, 50 and 80\% (bottom to top)
  Lagrangian radii (dotted curves). Red curve indicates a smoothed core radius
  using a method in \citet{2006MNRAS.368..677H}. Bottom panel: the binding
  energy of the hardest binary scaled by $f_{\rm max}$. 
  Red curves indicate the largest $E_{\rm bin}$ achieved in this simulations, and
  black dots indicates the largest $E_{\rm bin}$ at each time. 
  Blue dashed lines indicate the total energy of the cluster, $E$, and also 
  $0.5E$, and $(M_{\rm c}/M)E$ from top to bottom. In our model, $E = 1.5NkT$.
  The left panels give the results for the model with $W_0=3$ for equal-mass 
  particles,  and the right panels give the results for a mass function 
  with $\alpha = 2.35$ and $f_{\rm max}=8$.
\label{fig:core_collapse}}
\end{figure*}

In Figure \ref{fig:core_collapse} we present the evolution of the core
density and the core radius for an equal-mass model with $W_0=3$,
$N=$2k (left) and the same model but with a Salpeter mass function
with $f_{\rm max} = 8$ (right). 
We calculate the core radius and density using a method of 
\citet{1985ApJ...298...80C}, but we took into account the mass of the 
particles. Compared to the
equal mass case, the core collapse in the models with $f_{\rm max}=8$ 
is less clear, although the core
noticeably expands after a slight depression.

\begin{figure*}
\begin{center}
\includegraphics[width=80mm]{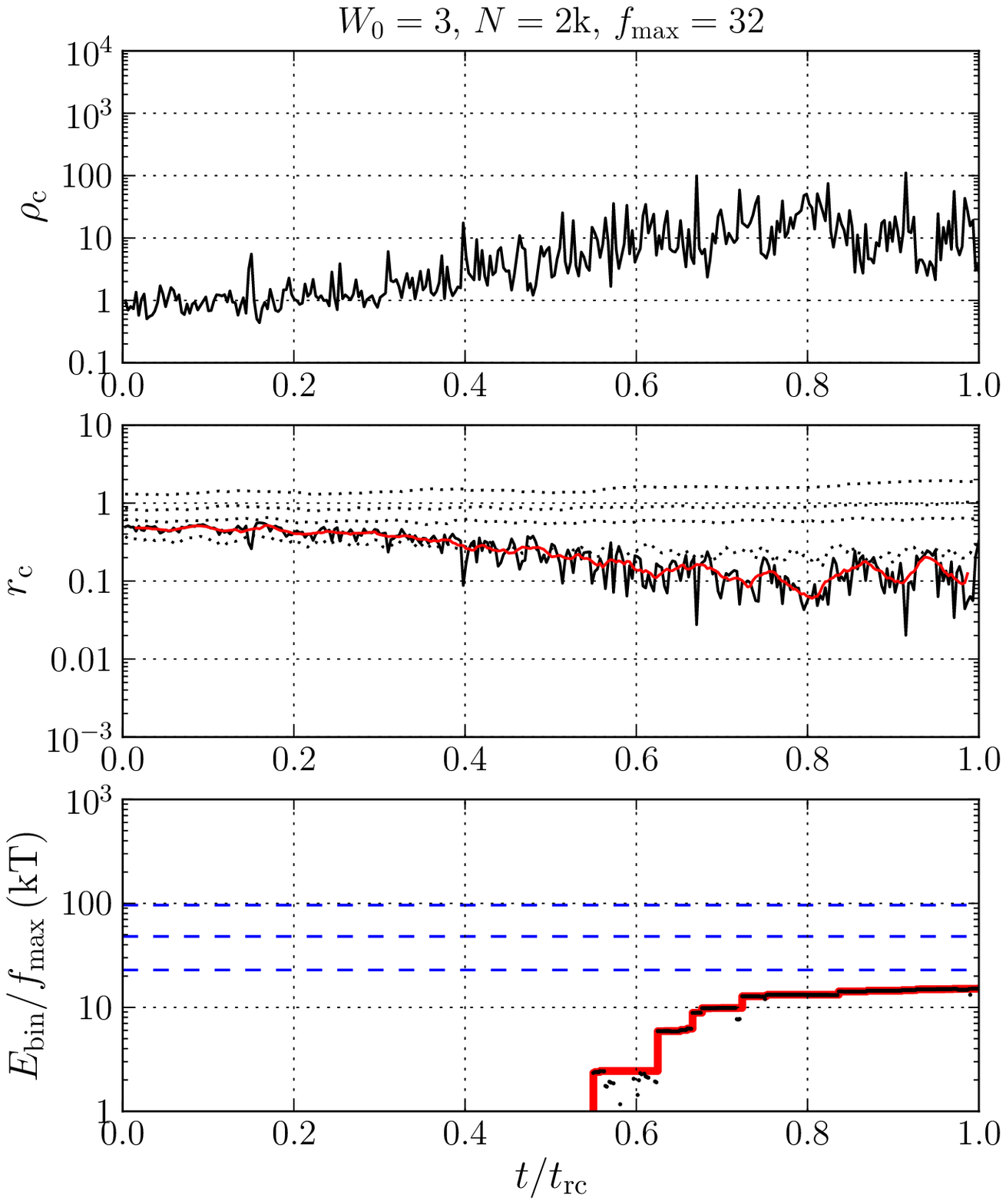}
\includegraphics[width=80mm]{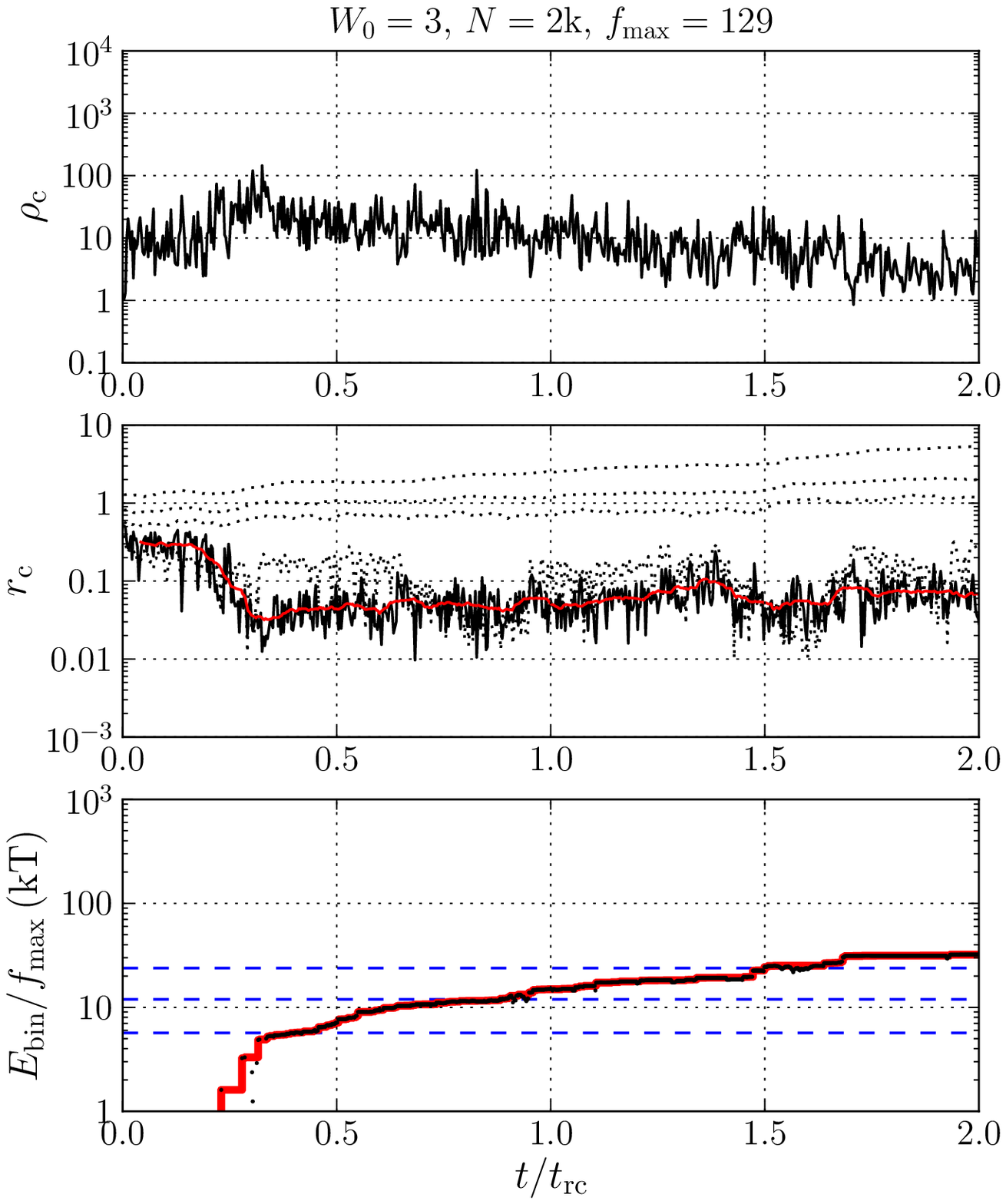}
\end{center}
\caption{Same as Figure \ref{fig:core_collapse}, but for models with
  $f_{\rm max}=32$ (left) and 129 (right).
 \label{fig:core_collapse1}}
\end{figure*}

If we increase $f_{\rm max}$, the core collapse
becomes more ambiguous.  In Figure \ref{fig:core_collapse1} we present 
the results of the same simulation as we presented in Figure
\ref{fig:core_collapse}, but with $f_{\rm max}=32$
(left in Fig.\,\ref{fig:core_collapse1}) and 129 (right).
After the core shrinks, it keeps the small core radius and
slowly expands (see right panel of Figure \ref{fig:core_collapse1}).

For models with equal-mass or small $f_{\rm max}$, we can easily measure 
the moment of core collapse and confirm the results. 
We tried two measurement methods to determine the moment of core collapse.
One is the moment when the core density
reached its maximum, and the other is the moment when the 
smoothed core radius (red curves in the middle panels of Figures 
\ref{fig:core_collapse}
and \ref{fig:core_collapse1}) reached its minimum 
\citep{2006MNRAS.368..677H} for models in which
the core evolution is visible ($f_{\rm max}\aplt 8$ for $N=2$k, 
$f_{\rm max}\aplt 32$ for $N=8$k, $f_{\rm max}\aplt 129$ for $N=32$k, and 
$f_{\rm max}\aplt 517$ for $N=128$k). 
We confirmed that there is no large discrepancy and no bias between them.

\begin{figure*}
\begin{center}
\includegraphics[width=80mm]{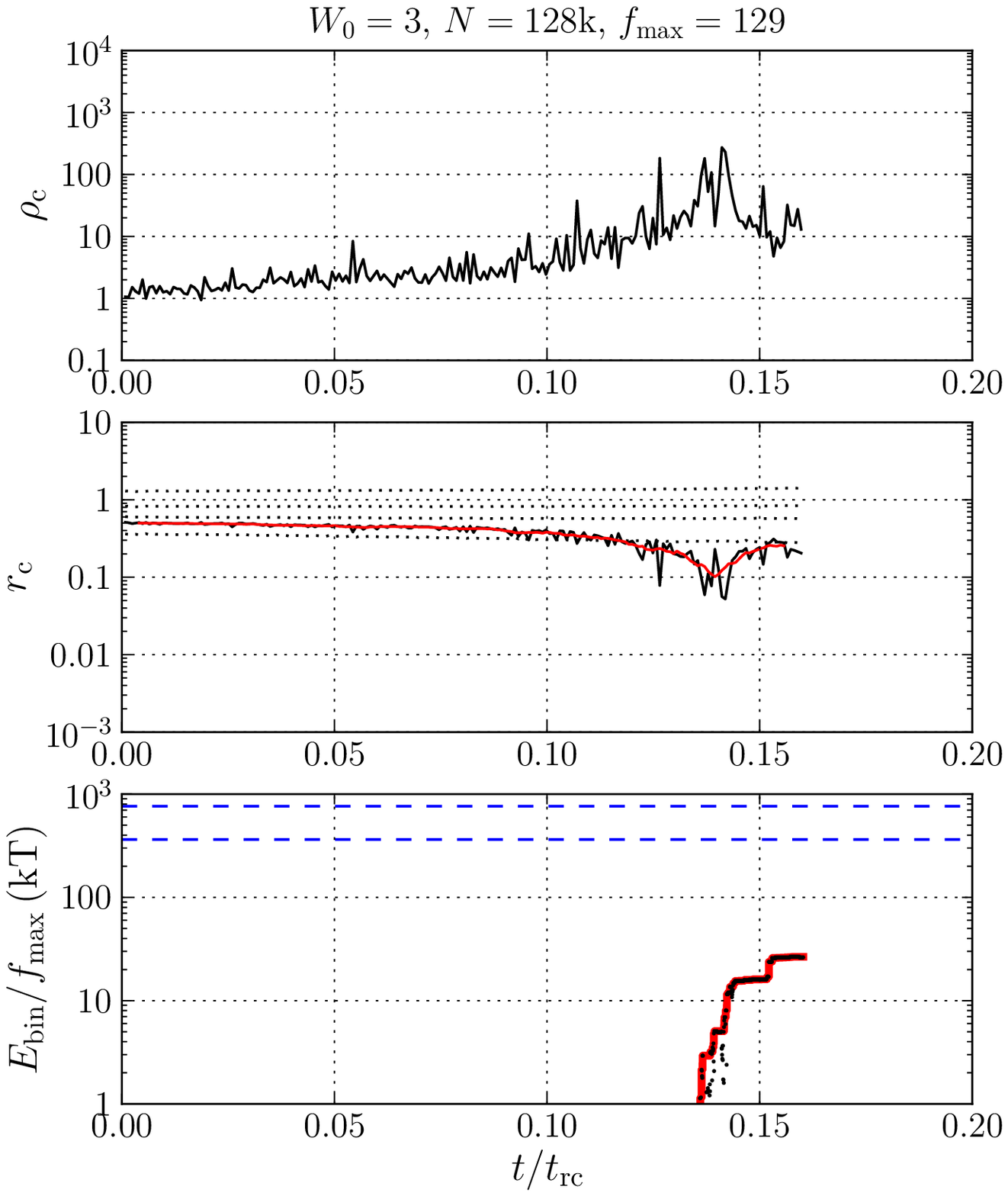}
\includegraphics[width=80mm]{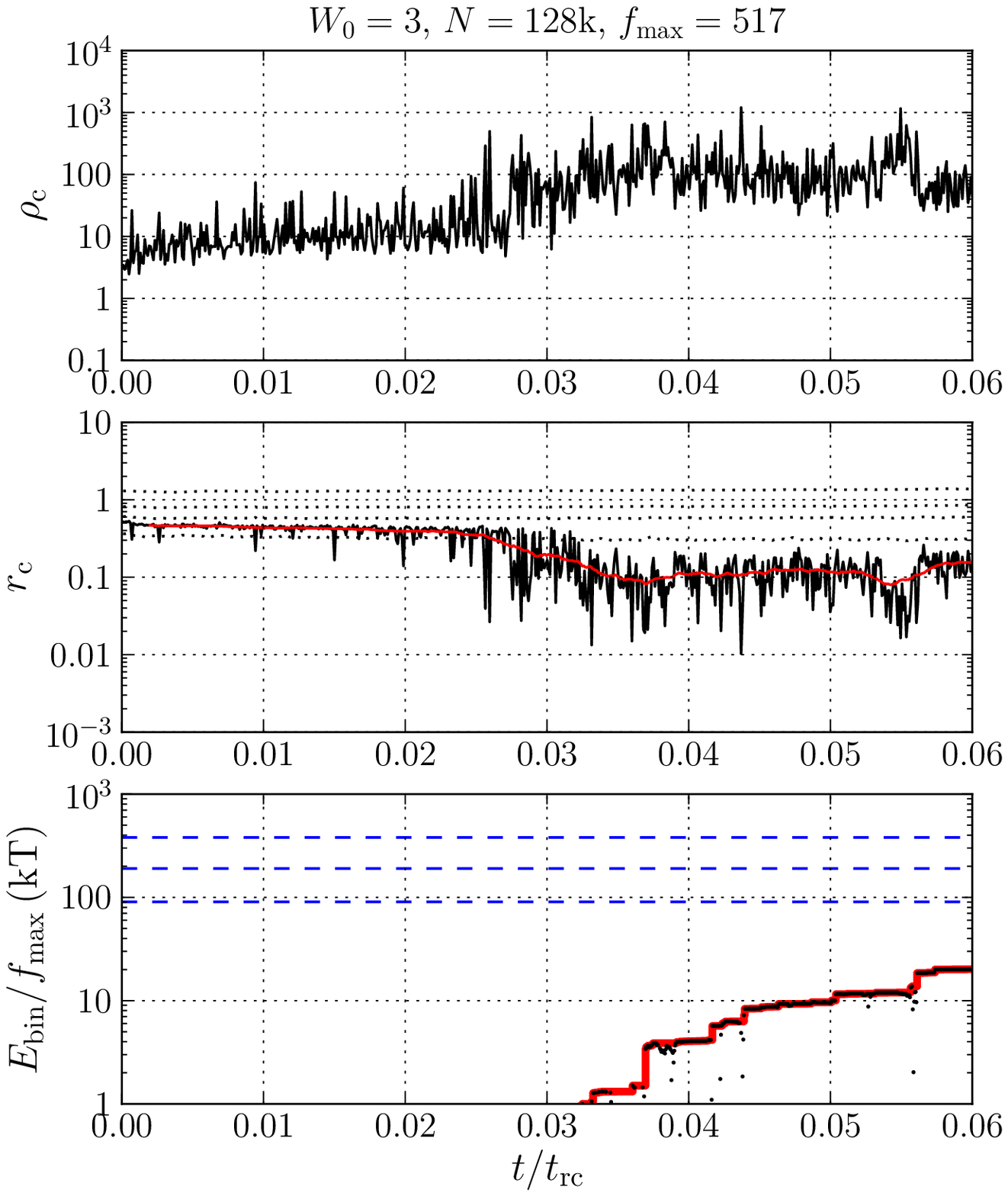}
\end{center}
\caption{Same as Figure \ref{fig:core_collapse}, but for models with
  $N=$128k and with $f_{\rm max}=129$ (left) and 517
  (right).\label{fig:core_collapse2}}
\end{figure*}

Based on our simulations we conclude that a clear core bounce occurs if
$M/m_{\rm max}\apgt 10^{3}$ ($f_{\rm max}\aplt 2$ for $N=2$k), and the
rapid core expansion after a shrink of the core becomes apparent only 
if $M/m_{\rm max}\apgt 100$
($f_{\rm max}\aplt 20$ for $N=2$k).  We illustrate these in the left
panels of 
Figure \ref{fig:core_collapse2}, where we present the same
models as in the right panels in Figure \ref{fig:core_collapse1} 
but with $N=128$k. This
model satisfies the first criterion ($M/m_{\rm max}\apgt 10^{3}$) and
as expected the core bounce is clearly visible. With $f_{\rm max}
=517$ (the right panels of Figure \ref{fig:core_collapse2}), however,
which does not satisfy either of the criteria and as a consequence the
core bounce becomes indistinguishable.  These criteria are similar to
the Spitzer instability \citep{1987degc.book.....S}. In the case of
multi-mass components, the criterion for the Spitzer instability is
$M/m_{\rm max}\sim 10^{4}$ \citep{2012MNRAS.425.2493B}.  We discuss
these criteria further in section 4.

\subsection{The determination of core-collapse time using hard binaries}

In the previous subsection we discussed the lack of a core bounce for
the case where $M/m_{\rm max}\aplt 10^{3}$. In particular if $M/m_{\rm
  max}\aplt 100$, the core in these cases however still expands quite
dramatically after some time. In these cases it becomes very hard to
use the core size or density peak to determine the moment of core
collapse, but the expansion of the core indicates that something like
a core collapse must have happened. In order to quantify this we focus
on the evolution of hard (dynamically formed) binaries, which are
suspected to generate the energy that causes the core to bounce.

\subsubsection{Measured binary hardness at core collapse}
In the bottom panels in Figures \ref{fig:core_collapse},
\ref{fig:core_collapse1}, and \ref{fig:core_collapse2}, we present the
binding energy of the hardest binaries in the various simulations. 
By comparison of the evolution of the binding energy with the
core density (or core radius), we observe that the moment of the core
bounce is consistently occurring at the same moment that the 
binding energy of the hardest binary reaches $\sim f_{\rm max} kT$.

\begin{figure*}
\begin{center}
\includegraphics[width=80mm]{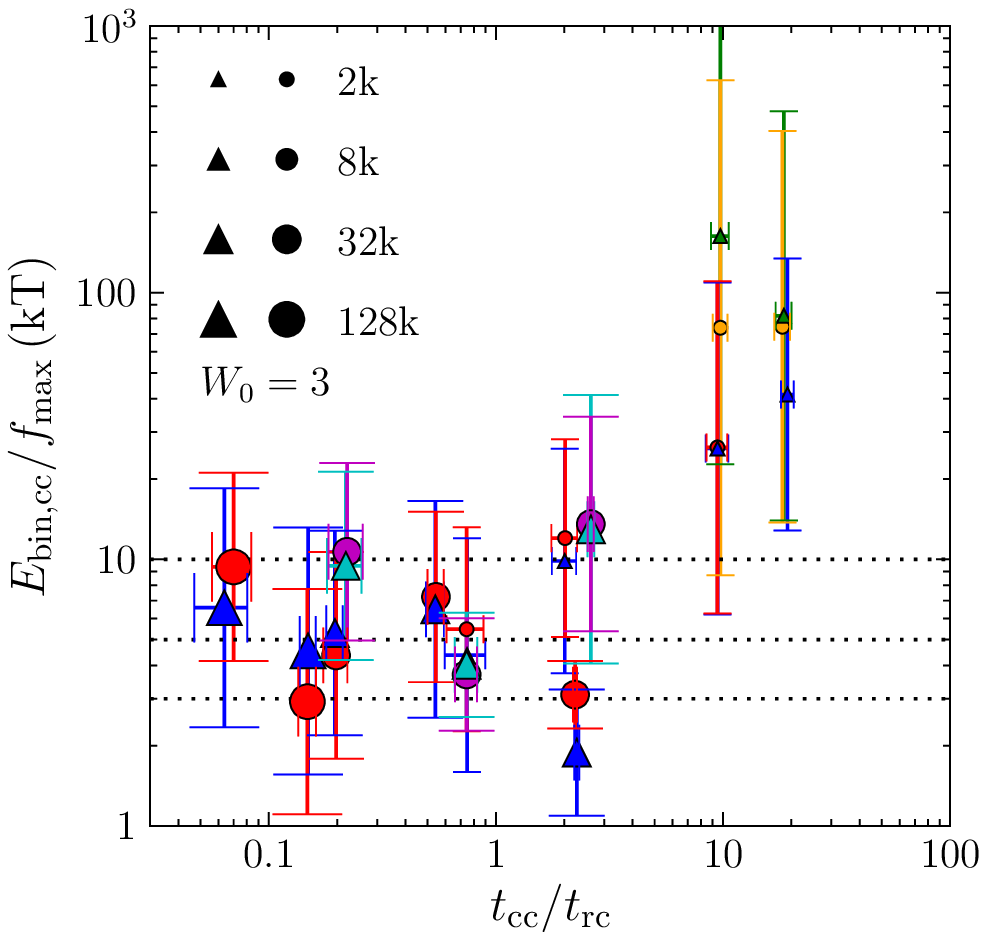}
\includegraphics[width=80mm]{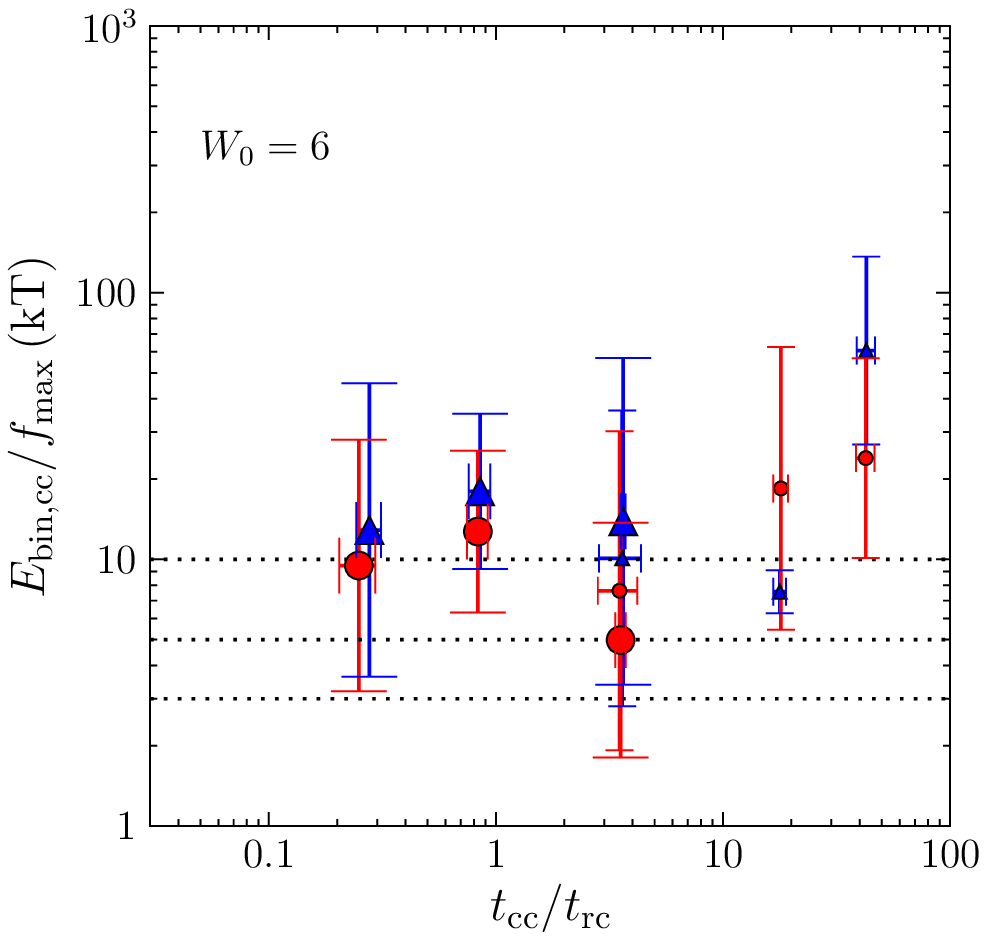}
\end{center}
\caption{The binding energy of the hardest binary at the moment of
  core collapse measured from time averaged core radius (circles)
  and from the maximum core density (triangles).  Blue and red indicate
  models with $\alpha=2.35$ without softening. Green and orange are for 
  the same models but with softening. Cyan and magenta indicate models 
  with $\alpha=1.2$ The binding energy is scale by $f_{\rm max}$.  Here we
  plot only models with $M/m_{\rm max}\apgt 100$ 
  ($f_{\rm max}\leq 8$ for $N=2$k, $f_{\rm max}\leq 32$ for $N=8$k, 
  $f_{\rm max}\leq 129$ for $N=32$k, and $f_{\rm max}\leq 517$ for $N=128$k) 
  and with $N_{\rm run}>1$.
\label{fig:tcc_kT_sc}}
\end{figure*}

In Figure \ref{fig:tcc_kT_sc} we present the binding energy of the
hardest binary at the moment of core collapse measured from the
highest core density and the smallest core radius ($E_{\rm bin, cc}$). 
In this figure we
scale the binding energy by a factor $f_{\rm max}$.  We measure the
moment of core collapse using two different methods, both of which
give consistent results. The scatter in the binding energy is larger,
probably because of the measurement timing. We can only measure the
binding energy at the moment of an output time. In particular for
equal-mass models and those with $f_{\rm max}=2$ the binding energy
rapidly increase towards the moment of core collapse (see Figure
\ref{fig:core_collapse}).  The softening in the gravitational
potential does not appear to affect our measurements of the moment of
core collapse, but $E_{\rm bin, cc}$ is systematically larger is the
softened models. We are therefore prone to overestimating the binding
energy in these models. 
We conclude that the average binding energy at the moment of
core collapse is $\sim 10f_{\rm max}kT$ and this effect appears to 
be independent of$N$. Hereafter we specify the
critical binding energy, $E_{\rm cr}$, as a minimum binding energy
required for core collapse.

\begin{figure*}
\begin{center}
\includegraphics[width=80mm]{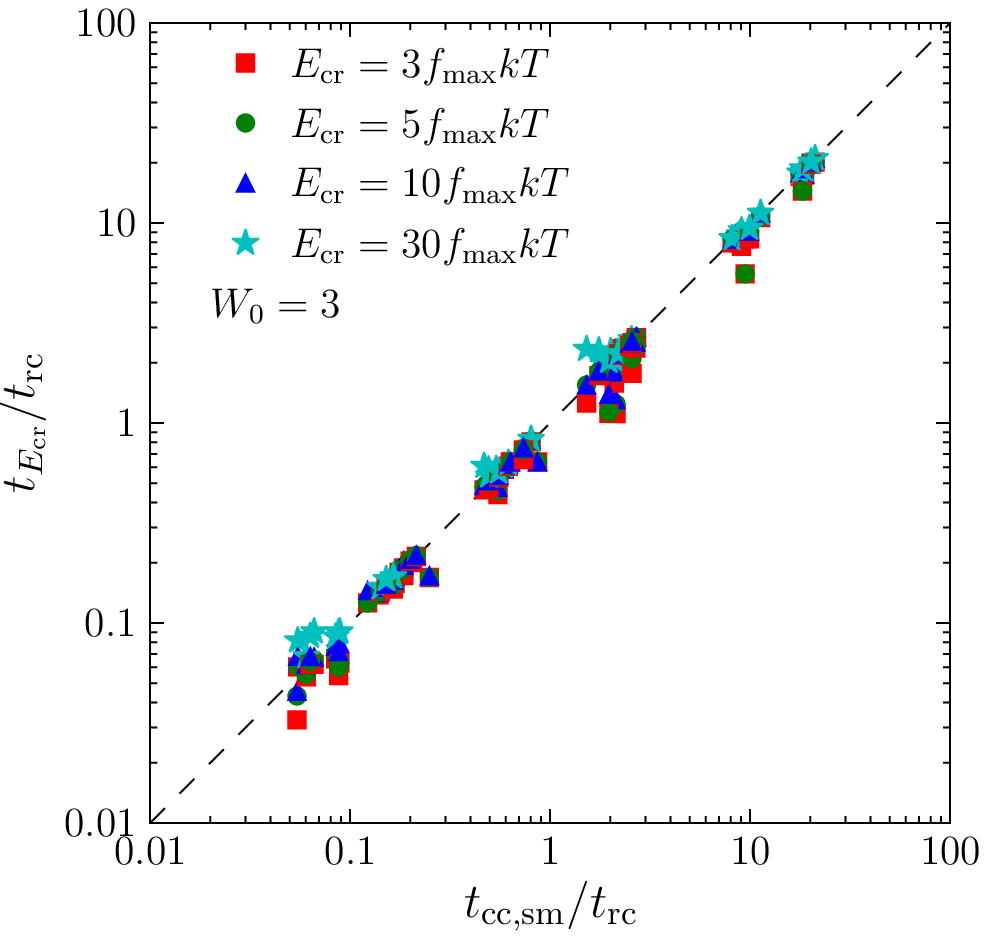}
\includegraphics[width=80mm]{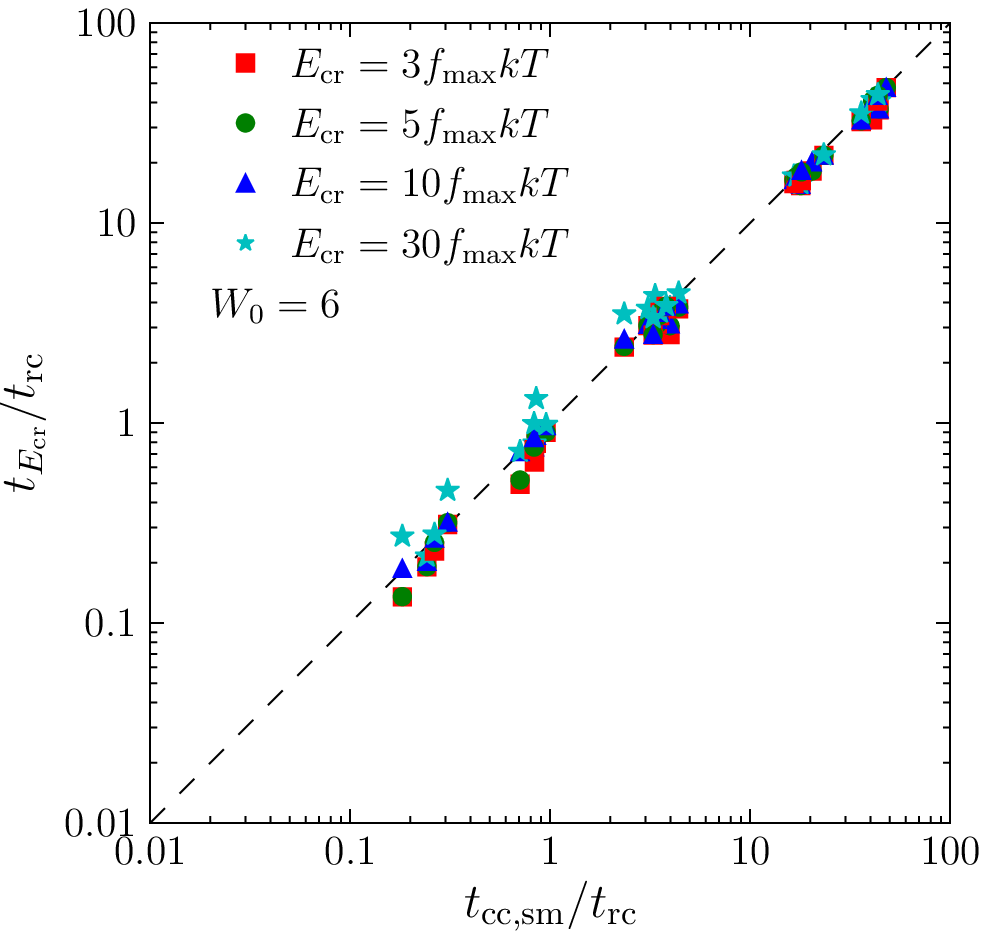}
\end{center}
\caption{Comparison of two independent measurements of the moment of
  core collapse.  
  Core-collapse time measured by smoothed core radius 
  and the time of the formation of $E_{\rm cr}$ binaries 
  for $W_0=3$ (left panel) and
  for $W_0=6$ (right panel). We adopted $E_{\rm ct}=3$, 5, 
  10, and 30$f_{\rm max} kT$ (red squares, green circles, blue triangles, and
  cyan stars, respectively). We plot only models with $M/m_{\rm max}\apgt 100$ 
  ($f_{\rm max}\leq 8$ for $N=2$k, $f_{\rm max}\leq 32$ for $N=8$k, $f_{\rm max}\leq 129$ for $N=32$k,
  and $f_{\rm max}\leq 517$ for $N=128$k).\label{fig:tcc_tkT}}
\end{figure*}

We measure the time when the binding energy of the hardest binary for
the first time reaches $E_{\rm cr}=$ 3, 5, 10, and $30f_{\rm max}kT$.
The results compared to the core collapse time measured from the
smoothed core radius are shown in Figure \ref{fig:tcc_tkT}. We find
that $E_{\rm cr}= 10f_{\rm max}kT$ provides the best comparison. In
the following analysis, we associate the first moment when the binding
energy of the hardest binary exceeds $10f_{\rm max}kT$ as the core
collapse time, even in the cases that the core collapse is not obvious 
upon the
inspection of the core radius (see Figure \ref{fig:core_collapse1} and
\ref{fig:core_collapse2}).  It turns out that the binding energy of
the hardest binary is an excellent indicator for identifying the
moment of core collapse.  In the following we discuss the
argument for $E_{\rm cr} \sim 10 f_{\rm max}kT$ from a more
theoretical perspective.
 
\subsubsection{Theoretical binary hardness at core collapse}

We estimate the critical binding energy for the core bounce from a
discussion on the energy emitted by the hard binary via a three-body
encounter.  In the dynamical evolution of star clusters through core
collapse, the cluster responds to a core collapse by a bounce, and the
occurrence is associated with the moment when the energy produced by
hard binaries exceeds the potential energy of the cluster core
$\phi_{0}$ \citep{1996IAUS..174..121H,2003gmbp.book.....H}.  Following
the discussion in \citet{1996IAUS..174..121H}, assuming that the core
is virialized until the moment of the bounce, the potential energy of
the core is
\begin{eqnarray}
|\phi_{0}| = N_{\rm c}\langle m \rangle _{\rm c}\sigma_{\rm c, 3D}^2,
\end{eqnarray}
where the $N_{\rm c}$, $\langle m \rangle _{\rm c}$, and $\sigma_{\rm
 c, 3D}$ are the number, the mean mass, and velocity dispersion of
the stars in the core, respectively. The energy released in an
encounter between a single star and a binary with binding energy
$E_{\rm bin}$ is estimated by $\Delta E_{\rm bin}=0.4E_{\rm bin}$
\citep{1975MNRAS.173..729H} for equal mass cases.  The coefficient is
ill constrained in multi-mass cases, and we therefore adopt $\Delta
E_{\rm bin} \sim E_{\rm bin}$ for the first order estimate of the
critical binding energy. We then obtain that
\begin{eqnarray}
E_{\rm cr} \sim N_{\rm c}\langle m \rangle _{\rm c} \sigma_{\rm c, 3D}^2.
\label{eq:E_cr}
\end{eqnarray}
With a mass function, $\langle m \rangle _{\rm c} > \langle m \rangle$
due to mass-segregation. Here we assume that $\langle m \rangle
_{\rm c} \sim m_{\rm max}$.  If we rewrite equation (\ref{eq:E_cr})
with $kT$, we obtain that
\begin{eqnarray}
E_{\rm cr} &\sim& N_{\rm c} \frac{m_{\rm max}}{\langle m \rangle} \frac{\sigma_{\rm c, 3D}^2}{\sigma_{\rm 1D}^2}\, kT\\
&\sim& 3N_{\rm c} f_{\rm max} \frac{\sigma_{\rm c, 1D}^2}{\sigma_{\rm 1D}^2}\, kT
\end{eqnarray}

Now we have to estimate $N_{\rm c}$ and $\sigma_{\rm c, 1D}^2$.
Initially $\sigma_{\rm c, 1D}^2 = 1.5$--1.6$\sigma_{\rm 1D}^2$ for
King models with $W_0=3$--6. The core velocity dispersion $\sigma_{\rm
  c, 1D}$ increases towards the core collapse
\citep{1996MNRAS.279.1037G}, but by only a factor of 2 because the
core evolves following $\rho_{\rm c}\propto r^{-\kappa}$ and $\kappa
=$ 2.2--2.3
\citep{1980ApJ...242..765C,1980MNRAS.191..483L,1995PASJ...47..561T,
  2003gmbp.book.....H}. We therefore estimate that $\sigma_{\rm c,
  1D}^2/\sigma_{\rm 1D}^2$ is roughly a factor of 3.  For $N_{\rm c}$
it is theoretically estimated that $N_{\rm c} \sim 80$
\citep{1996IAUS..174..121H}.  Numerically it is obtained that $N_{\rm
  c}=$10--30 for an equal-mass system \citep{1996ApJ...471..796M},
$N_{\rm c}=10$--100 for two-component systems
\citep{2007MNRAS.374..703K}, and $N_{\rm c}\sim 25$ for multi-mass
systems \citep{2003gmbp.book.....H}.  Recent study by
\citep{2012NewA...17..272T} report that there are only $\sim 5$ stars
in the core when a hard binary with $\sim 10kT$ forms and that the
formation process of such a hard binary is sudden rather than gradual
evolution from a softer binary. From our numerical result that $E_{\rm
  cr}\simeq 10f_{\rm max}kT$ and analytical estimation that $E_{\rm
  cr} \sim 9 N_{\rm c}f_{\rm max}kT$ we roughly estimate that $N_{\rm
  c}\sim O(1)$. Hereafter we adopt $E_{\rm bin} > E_{\rm cr} = 10f_{\rm
  max}kT$ as the moment of core collapse.

\subsection{The core-collapse time} 

\begin{figure*}
\begin{center}
\includegraphics[width=80mm]{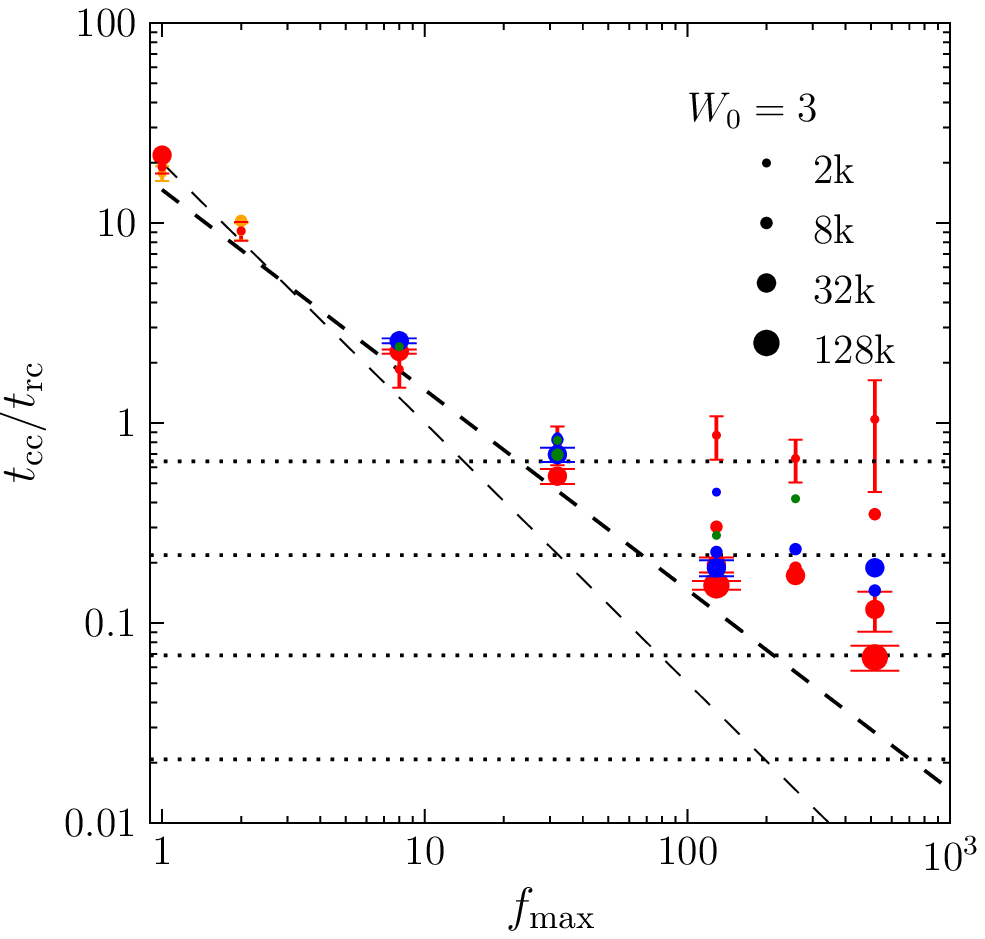}
\includegraphics[width=80mm]{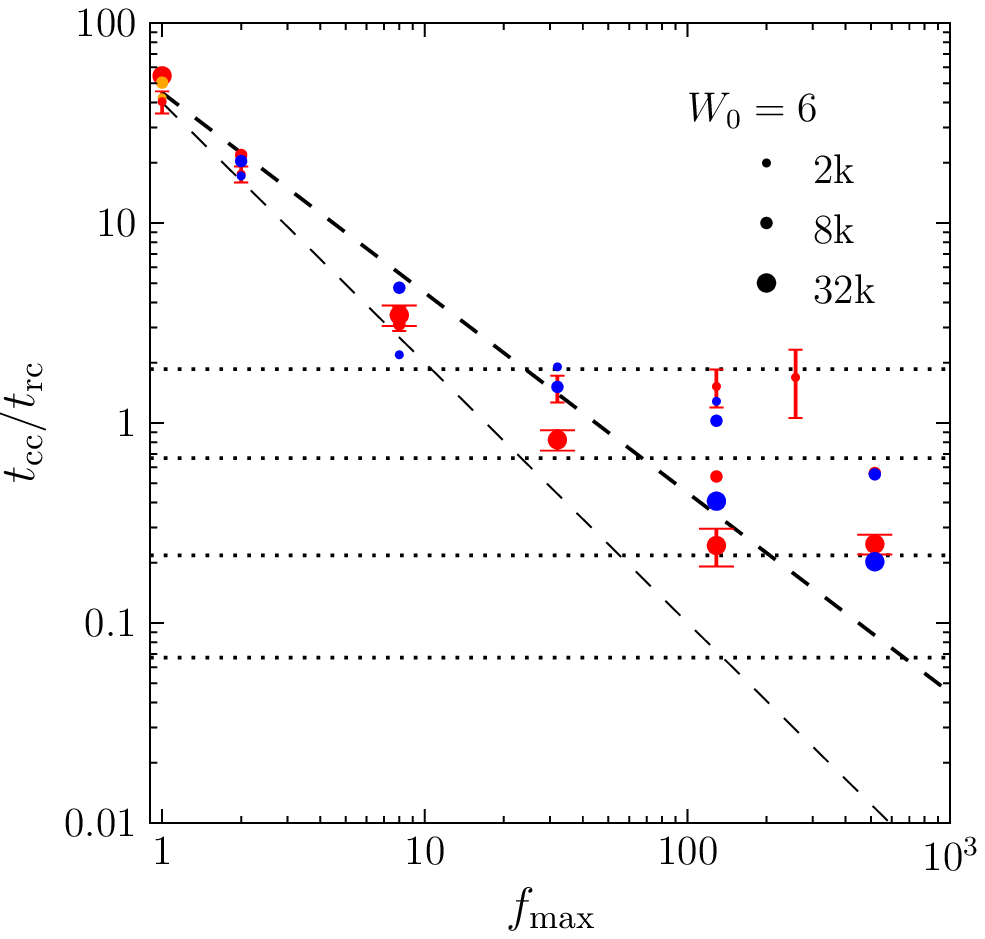}
\end{center}
\caption{The core-collapse time as a function of the maximum mass of
  cluster particles, $f_{\rm max}$ for models with
  $W_0=3$ (left panel) and 6 (right panel).  Red, blue, and green
  indicate models with $\alpha=2.35$ (Salpeter), 1.7, and 1.2,
  respectively. Orange indicate models with a softening length. 
  The sizes of the symbols indicate the number of
  particles. Thick dashed line shows the analytic results obtained from
  equation (\ref{eq:df2}) and thin dashed line indicates 
  $\propto f_{\rm max}^{-1.3}$. The dotted lines indicate the minimum
  core-collapse time obtained from $10t_{\rm cross}$. They are for
  $N=$ 2k, 8k, 32k, and 128k from top to bottom.
\label{fig:t_cc_w3}}
\end{figure*}

In Figure \ref{fig:t_cc_w3} we present $t_{\rm cc}/t_{\rm rc}$ as a
function of $f_{\rm max} (=m_{\rm max}/\langle m \rangle)$. Here we
defined the core-collapse time $t_{\rm cc}$ as the moment when the
binding energy of the hardest binary in the cluster exceeds $E_{\rm
  cr}=10f_{\rm max}kT$.  The core-collapse time for single-mass
component models is $t_{\rm cc}/t_{\rm rc} \simeq 20$ for $W_0=3$ and
$50$ for $W_0=6$, which are consistent with previous results
\citep[][and references therein]{2004ApJ...604..632G}.  With a mass
function, $t_{\rm cc}/t_{\rm rc}$ decreases as we increase $f_{\rm
  max}$, and it follows $f_{\rm max}^{-1}$ (thick dashed line in
Figure \ref{fig:t_cc_w3}) as far as $f_{\rm max} \aplt 30$.

We analytically derive the core-collapse time for models with stellar
mass functions using the dynamical friction timescale of the most
massive stars in the cluster, assuming that star clusters collapse
when the most massive stars reach the cluster center. The dynamical
friction timescale of the most massive star with $m_{\rm max}$ is
estimated from a simple equation. We follow the description in
\citep{2002ApJ...576..899P} \citep[see also section 8.1.1
  in][]{2008gady.book.....B}, in which the dynamical friction
timescale of a black hole which spirals in to the galactic center is
derived.  We assume that the massive star has initially a circular
orbit with velocity, $v_{\rm c}$, at a distance $r$ from the cluster
center.  From equation (8.9) in \citet{2008gady.book.....B}, the
frictional force $F=m_{\rm max}|dv_{m}/dt|$ on the massive star is
\begin{eqnarray}
F = \frac{4\pi G^2 m_{\rm max}^2 \rho (r) \ln \Lambda '}{v_{\rm c}^2} 
\left[\mathrm{erf}(X)- \frac{2X}{\sqrt{\pi}} \exp(-X^2) \right ],
\label{eq:F}
\end{eqnarray}
where $X \equiv v_{\rm c}/(\sqrt{2}\sigma_{\rm 1D}) = 1$, where
$\sigma_{\rm 1D}$ is the one-dimensional velocity dispersion and
$v_{\rm c}$ is equivalent to the two-dimensional velocity dispersion.
The value of the Coulomb logarithm here is different from that in
equation (\ref{eq:tcrlx}), and therefore we write $\ln \Lambda '$.  
In order to simplify this equation, we
assume the density distribution to be a singular isothermal sphere,
$\rho (r) = v_{\rm c}^2/(4\pi G r^2)$, and equation (\ref{eq:F})
then becomes
\begin{eqnarray}
F = 0.428  \ln \Lambda ' \frac{G m_{\rm max}^2}{r^2}.
\end{eqnarray}
The angular momentum change of 
the massive star due to the friction is
\begin{eqnarray}
\frac{dL}{dt} &=& -Fr\\
&\simeq& 0.428 \ln \Lambda ' \frac{G m_{\rm max}^2}{r}.
\label{eq:dLdt}
\end{eqnarray}
In an isothermal sphere the circular velocity is independent of
radius, and the angular momentum at radius $r$ is written as
$L=m_{\rm max}rv_{\rm c}$.  With equation (\ref{eq:dLdt}), we obtain
\begin{eqnarray}
r\frac{dr}{dt} = -0.428\ln \Lambda '  \frac{Gm_{\rm max}}{v_{\rm c}}.
\end{eqnarray}
The dynamical friction timescale of the most massive star is finally
written as 
\begin{eqnarray}
t_{\rm df} = \frac{1.91}{\ln \Lambda '}\frac{r^2\sigma_{\rm 3D}}{G m_{\rm max}}.
\label{eq:df}
\end{eqnarray}
Assuming that $t_{\rm df}=t_{\rm cc}$, from equations (\ref{eq:tcrlx}) and 
(\ref{eq:df}) we obtain 
\begin{eqnarray}
\frac{t_{\rm cc}}{t_{\rm rc}} = 29.4 \frac{\ln \Lambda}{\ln \Lambda'} 
f_{\rm max}^{-1} \frac{G r^2 \rho_{\rm c} \sigma _{\rm 3D}}{\sigma _{\rm c, 3D}^3}.
\label{eq:df2}
\end{eqnarray}
This result is shown in Figure \ref{fig:t_cc_w3} (dashed curves) and
is consistent with the simulations. Here, we adopted $r$ to be the
virial radius, $r_{\rm vir}=GM/4|E|$=1 in $N$-body units, and $\Lambda
' = 0.1 N$ \citep{1994MNRAS.268..257G,2003gmbp.book.....H}.  For the
Coulomb logarithm for $t_{\rm rc}$ (equation (\ref{eq:tcrlx})), we
adopted $\Lambda = 0.1 N_{\rm c}$, where $N_{\rm c}\equiv (M_{\rm c}/M)N$
is the number of particles in the core.  The dynamical friction
timescale of the most massive star is proportional to $m_{\rm
  max}^{-1}$, and as a consequence the core-relaxation time is
proportional to $\langle m \rangle ^{-1}$ (see Eq.\,\ref{eq:tcrlx}),
and therefore we obtain that $t_{\rm cc}/t_{\rm rc} \propto (m_{\rm
  max}/\langle m \rangle)^{-1}$.

\subsection{The minimum core-collapse time}

We see in Figure \ref{fig:t_cc_w3} that for models with fewer
particles ($N=2$k) the value of $t_{\rm cc}/t_{\rm rc}$ starts to
deviate from the analytic result for $f_{\rm max} \sim 30$.  This
critical value of $f_{\rm max}$ however, depends on $N$; for models
with $N=128$k the simulations and theory give consistent results up to
$f_{\rm max} \sim 100$. The models between $N=2$k and 128k show a
consistent picture in the sense that the models with a larger $N$
start to deviate from the theory at a larger value of $f_{\rm max}$.
We bolster our earlier conclusion that a core bounce requires that
$M/m_{\rm max} \apgt 100$. The core-collapse time saturates for a
smaller value of $f_{\rm max}$ in models with fewer particles. We will
discuss this critical value of $M/m_{\rm max}$ in section
\ref{discussion} and now focus on estimating the ``minimum''
core-collapse time, as indicated in Figure \ref{fig:t_cc_w3}.

This minimum in the core-collapse time depends on $N$ due to the
dependency of $t_{\rm rc}$ on $N$. We consider that this minimum
core-collapse time depends on the crossing time, $t_{\rm cross}$, of
the cluster, because the dynamical friction time cannot be shorter
than the crossing time. We adopt $t_{\rm cross}=r_{\rm
  vir}/\sigma_{\rm 1D}$, where $r_{\rm vir}$ is the virial radius.
The minimum core-collapse time obtained from the simulations is
roughly consistent with $10t_{\rm cross}$.  The dotted curves in
Figure \ref{fig:t_cc_w3} give $10t_{\rm cross}$, and they depend on
$N$, because $t_{\rm cross}$ is independent of $N$ whereas $t_{\rm
  rc}$ is.

\subsection{The maximum critical binding energy}

\begin{figure*}
\begin{center}
\includegraphics[width=80mm]{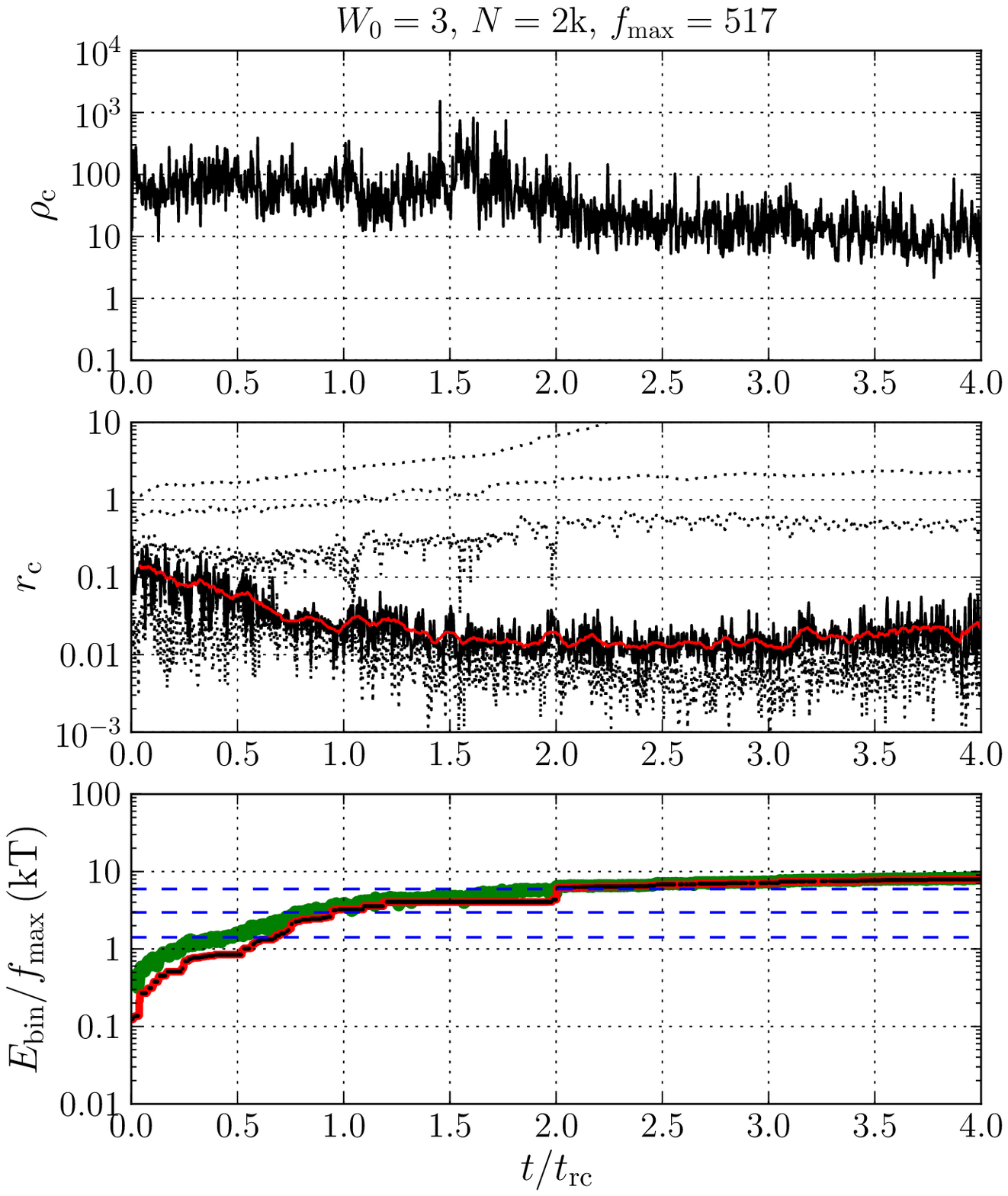}
\includegraphics[width=80mm]{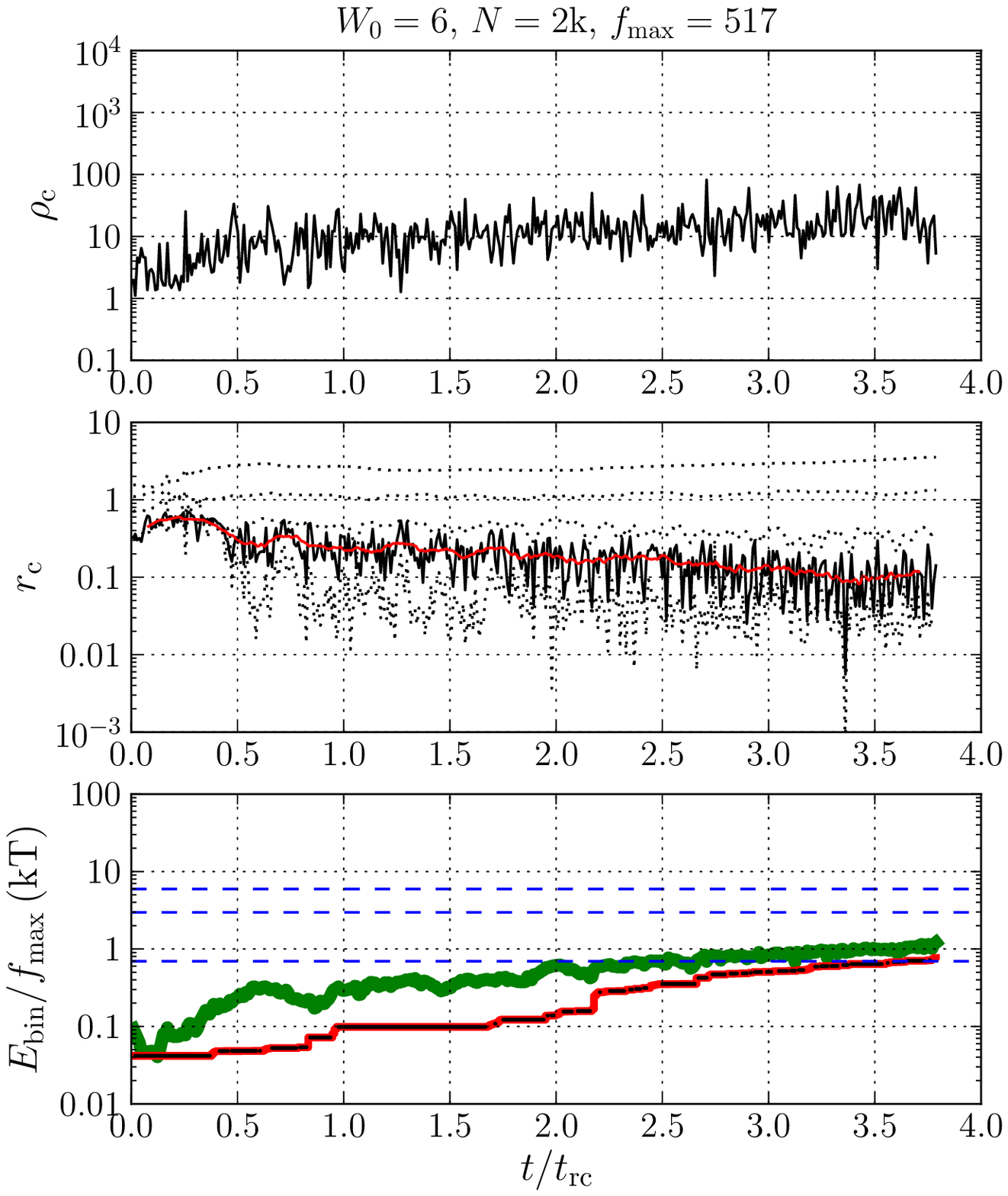}
\end{center}
\caption{Same as Figure \ref{fig:core_collapse}, but for models with
  $N=$2k and $f_{\rm max}=517$ for $W_{0}=3$ (left) and $W_{0}=6$
  (right). In bottom panels, green curves show the total binding energy of 
binaries.
\label{fig:core_collapse3}}
\end{figure*}

For models with a large $f_{\rm max}$ the critical binding energy
$E_{\rm cr}$ is comparable to the total energy of the cluster ($E$).
In those models, for example $W_0=3$, $N=2$k, and $f_{\rm max}=517$, 
the cluster dissolves before the binding energy reaches $10f_{\rm
  max}kT$. We find that $E_{\rm cr}=0.5E$ roughly traces the minimum
of the smoothed core radius (see left panels of Figure
\ref{fig:core_collapse3}).  We therefore adopt $E_{\rm cr}=0.5E$ if
$10f_{\rm max}kT>0.5E$. We are able to detect the moment of core-collapse 
time even for $m_{\rm max} \sim M_{\rm c}$ if we adopt $E_{\rm cr} = 0.5E$ 
(see right panels of Figure \ref{fig:core_collapse3}). 

When $m_{\rm max} > M_{\rm c}$, the
evolution of the hardest binary is different from those in models with
$m_{\rm max}<M_{\rm c}$.  In the former case, a hard sub-systems in which 
several stars orbits around the most massive star is actually detected. 
The hardest binary in
this sub-system gradually hardens due to repeated scattering
encounters with other stars.  This evolution is visible in the
temporal behavior of the total binding energy of the binaries. In
Figure \ref{fig:core_collapse3} (right panels), we present the total
binding energy (green curve), which represents the total energy of the
sub-system with the most massive star. The binding energy of the
hardest binary is initially much smaller than the total binding
energy, but it eventually catches up with the total binding
energy. When on the other hand $m_{\rm max}\aplt M_{\rm c}$, the total
binding energy remains comparable to the binding energy of the hardest
binary (Figure \ref{fig:core_collapse3}, left panels).  In both cases,
the binding energy seems to be limited by the total energy of the
cluster. In our simulations, the binding energy evolution saturates
between $E$ and $(M_{\rm c}/M)E$ (see also Figures \ref{fig:core_collapse}
and \ref{fig:core_collapse1}).  In table \ref{tb:E_cr} we summarize
the adopted values of the critical binding energy $E_{\rm cr}$.

\begin{table*}
\begin{center}
\caption{Summary of $E_{\rm cr} $\label{tb:E_cr}}
\begin{tabular}{ccccc}\hline\hline
$m_{\rm max}$  & $E_{\rm cr}$ &  $M/m_{\rm max} $ & \multicolumn{2}{c}{$f_{\rm max}$ in our models ($N=2$k)}\\
           &            &    $(\equiv N_{\rm eff})$     & $W_{\rm 0}=3$ & $W_{\rm 0}=6$ \\
 \hline
 $m_{\rm max}<M_{\rm c}$  & $10f_{\rm max}kT$  &  $\apgt10$ &  $<258$ (Fig. \ref{fig:core_collapse}, \ref{fig:core_collapse1})& $<129$ \\
 $m_{\rm max}\sim M_{\rm c}$ & $0.5E$ & $\sim 10$        & 258, 517 (Fig. \ref{fig:core_collapse3}) & 129, 258 \\
 $m_{\rm max}>  M_{\rm c}$  & - (no collapse) & $\aplt10$  &  - & 517 (Fig. \ref{fig:core_collapse3})\\
\hline
\end{tabular}
\end{center}
\end{table*}

\section{Discussion}

\subsection{$N$-dependence and comparison with single-component models}
\label{discussion}

We aim to find an objective criterion for detecting the core collapse
in simulated star clusters.  We concentrate on those cases where
$M/m_{\rm max} \apgt 10^3$ and for rapid core expansion in the case
that $M/m_{\rm max} \apgt 100$. Here we will make an analogy with
single-component models.  From a wide range of analytic calculations
and simulations, it is well established that the dynamical evolution
of star clusters such as relaxation, core collapse, core bounce, and
gravothermal oscillations only depend the total number of particles in
the system (and in the core). For example, gravothermal oscillation
occurs only when the number of particles exceed $\sim 10^4$
\citep{1987ApJ...313..576G,1996ApJ...471..796M}.  This criterion comes
from the number of particles in the core after core bounce, $N_{\rm
  c}$.  (Here we define $N_{\rm c}$ as the average number of particles
in the core after the actual core collapse, and we adopt $N_{\rm cb}$
as the number of particles at the moment of core bounce, i.e; at the
moment of deepest core collapse.) The gravothermal oscillation occurs
only when $N_{\rm c}>N_{\rm cb}$. While the value of $N_{\rm cb}$ is
considered a constant in the range of 10 to 80
\citep{1987ApJ...313..576G,1996ApJ...471..796M,1996IAUS..174..121H},
$N_{\rm c}$ depends on $N$. If we adopt $N_{\rm c} \simeq N^{1/2}$
\citep{1996ApJ...471..796M}, we can confirm that $N>10^4$ satisfies
$N_{\rm c}>N_{\rm cb}$.

The behavior after core collapse changes when $N$ decreases.  So long
as $N_{\rm c} \simeq N_{\rm cb}$ (i.e., $10^3 \aplt N\aplt 10^4$) we
observe similar evolution but the collapse becomes shallower for
smaller $N$, and the gravothermal oscillations damp \citep[see Fig. 1
  in ][]{1996ApJ...471..796M}.  For $100 \aplt N\aplt 10^3$ the core
bounce becomes indiscernible \citep[see Figure 10 in
][]{1994MNRAS.270..298G}.  This transition is quite similar to those
we observe if $m_{\rm max}$ is increased.  Our results of the
multi-mass case are scalable to those of the single mass case if we
define an effective number of particles ($N_{\rm eff}$) in the latter.
For an equal mass system, $N_{\rm eff} = N$, but when we introduce a
spectrum of masses, $N_{\rm eff} = M/m_{\rm max}$.  Interestingly, a
similar conclusion is obtained from recent results for two-component
and multi-component systems
\citep{2012MNRAS.420..309B,2012MNRAS.425.2493B}.

\begin{figure}
\begin{center}
\includegraphics[width=80mm]{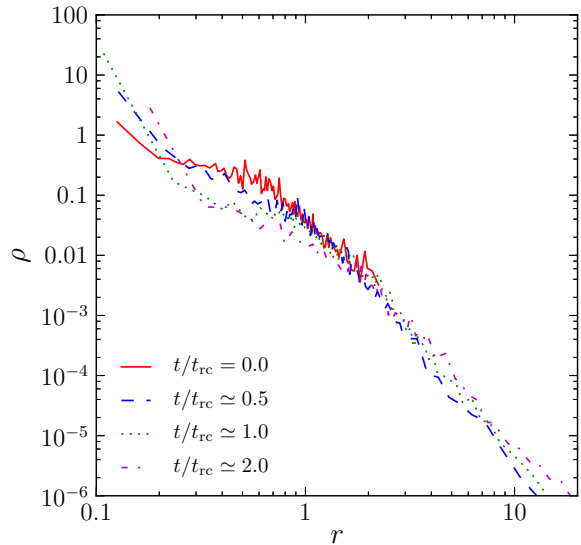}
\end{center}
\caption{The evolution of the density profile for one of model w3-2k-m129-Sal
(the same model in the right panel of Figure \ref{fig:core_collapse1}).
\label{fig:density_prof_ev}}
\end{figure}

If the number of particles drops below $\sim 100$, the behavior of
the $N$-body system changes from being a {\em many-body} system to a
{\em few-body} system, which evolves chaotically
\citep{1988ASSL..140..313M} rather than deterministic. In few-body
systems it is hard to notice the collapse of the core in the evolution
of core density and radius. In these cases the binary cannot harden to
$\apgt 100kT$ because the total energy budget of the cluster $\aplt
100kT$.  In such systems the hard binary stops interacting with other
cluster members when it becomes too tight and the surrounding density
becomes too low. As a result, the ``tenured'' or sometimes called
``frozen binary'' remains in the cluster
\citep{1985IAUS..113..305C}. This can be observed in Figures
\ref{fig:core_collapse1} and \ref{fig:core_collapse3}, for the 
multi-mass models when $M/m_{\rm
  max} \aplt 100$.  The dynamical evolution of tenured binaries almost
stops in this case, but they remain in the cluster (most likely in the
core). In Figure \ref{fig:density_prof_ev} we present the evolution of
density profiles of model w3-2k-m129-Sal (we used the same model in
the right panel of Figure \ref{fig:core_collapse1}).  The tenured
binary and its relatively low-density environment are noticeable as
the high density peak in the center and a dimple at around $0.2\,r$ in
the more extended core. This effect is similar to the core
mass-deficiency arguments used in galactic nuclei with binary black holes
after a major galaxy merger \citep{2010ApJ...718..739M}.  In the
formation process of tenured binaries, massive particles concentrated
in the cluster center are ejected from the cluster by sling-shot
interactions with the binary.  This mechanism leads to the formation
of massive runaway stars around dense, young star clusters
\citep{2011Sci...334.1380F}.

\subsection{Dynamical evolution driven by the most massive stars}

The dynamical evolution of equal-mass models scale with $t_{\rm rc}$
irrespective of the particle number \citep{1987degc.book.....S}.  Our
results are consistent with this hypothesis in the case of multi-mass
simulations, so long as $M/m_{\rm max}\apgt 100$.  For multi-mass
models, however, $t_{\rm cc}/t_{\rm rc}$ decreases for increasing
$m_{\rm max}/\langle m \rangle (= f_{\rm max})$, and we demonstrate
that $t_{\rm cc}$ is determined by the dynamical friction timescale of
the most massive stars in the cluster.  Here we demonstrate that the
dynamical evolution of star clusters with a mass function is driven by
the most massive stars in the cluster.

\begin{figure}
\begin{center}
\includegraphics[width=80mm]{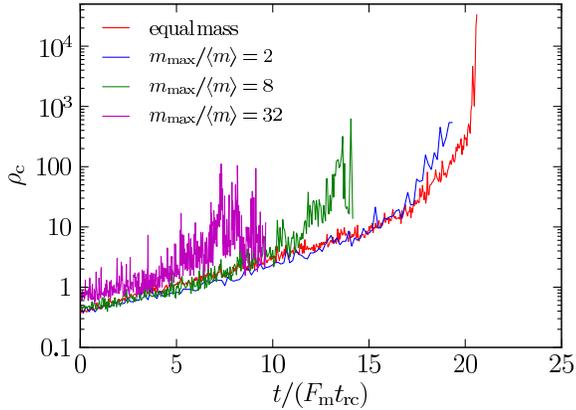}
\end{center}
\caption{Evolution of the core density scaled by $t_{\rm rc}$ and the
  factor of the relaxation time for the most massive stars, $F_{\rm
    m}$, for models with $N=8$k, $W_0=3$, and $\alpha =
  2.35$.\label{fig:ev_com_8k_w3}}
\end{figure}

The relaxation of a multi-mass system is dominated by the dynamics of
the most massive star, and the global relaxation time is a factor of
$F_{\rm m} \equiv \ln(\gamma N/f_{\rm max})/(f_{\rm max}\ln(\gamma
N))$ shorter than that of an equal mass system.  In Figure
\ref{fig:ev_com_8k_w3}, we present the evolution of the core density
scaled in time by the product of $t_{\rm rc}$ and $F_{\rm m}$.  So
long as the model satisfies the condition of core collapse ($M/m_{\rm
  max}\apgt10^{3}$), the evolutionary tracks of the core density are
scaled with $F_{\rm m}t_{\rm rc}$.  The scaling parameters are
determined using the initial cluster realization.  When the cluster
core starts to collapse, the models with a mass function start to
deviate from the equal-mass case.

\begin{figure}
\begin{center}
\includegraphics[width=80mm]{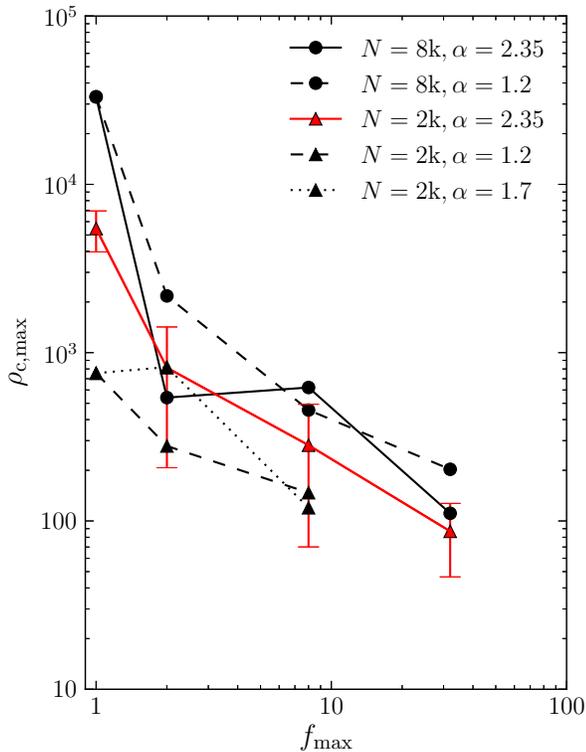}
\end{center}
\caption{The maximum density reached during core collapse as a
  function of the mass of the most massive star in terms of the mean
  mass, $f_{\rm max}$.
\label{fig:d_max}}
\end{figure}

We also observe the maximum core density, $\rho_{\rm c, max}$,
which is the core density at the core collapse time,
depends on $f_{\rm max}$. The maximum core density decreases 
when $f_{\rm max}$ increases.  We
present the relation between $\rho_{\rm c, max}$ and $
f_{\rm max}$ in Figure \ref{fig:d_max}.  This phenomenon
is similar to the relation between the maximum density and $N$
\citep{1994MNRAS.270..298G}.  \citet{1996IAUS..174..121H} derived that
$\rho \propto N^{-2}$, in which case we expect that the maximum
density decreases $\propto f_{\rm max}^{-2}$. Here
we assumed that the core consists of the most massive stars and
therefore that $\rho_{\rm c, max} \propto N_{\rm eff}^{-2}$.  In Figure
\ref{fig:d_max}, however, the power appears to be shallower than $-2$,
although the trend that the maximum density decreases for increasing
$f_{\rm max}$ is reproduced.  This might be caused
by the mean particle mass in the core being smaller than $m_{\rm
  max}$.  Here we would like to point out that we measure the core
density only in a snapshot, the moment of which is limited by our
output frequency.  We therefore are likely to miss the highest density
peak; it is very difficult to catch the exact moment of the highest
density in an $N$-body simulations. This can be solved by storing the
particle position and velocity information in a time resolved data
format, as was proposed by \citet{2010MNRAS.401.1898F} (see also
\citet{2012NewA...17..520F}).

\subsection{Comparison with previous results}

\begin{figure}
\begin{center}
\includegraphics[width=80mm]{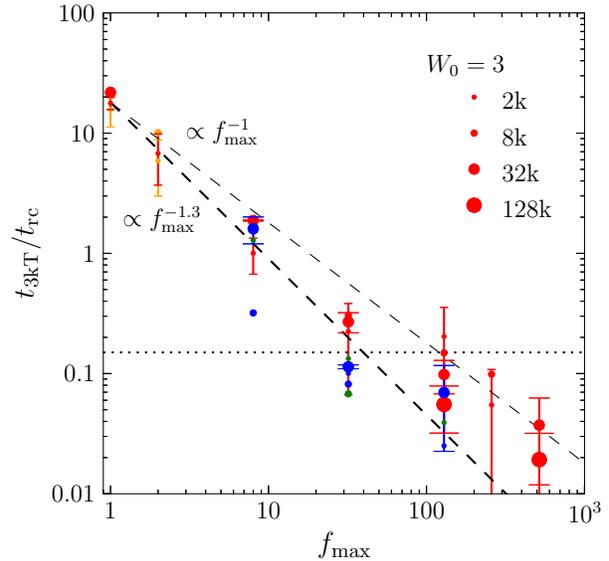}
\end{center}
\caption{The moment of the formation of the first 3kT binary
  \citep[the hard binary limit according to][]{2009PASJ...61..721T} as
  a function of the maximum mass of the cluster particles, $f_{\rm
    max}$ for $W_0=3$ models. Colors are the same as 
  Figure \ref{fig:t_cc_w3}.
  Models with $N=2$k and $f_{\rm max} = 517$ initially host 3kT
  binaries, and therefore they are not plotted in this figure.  Thick and thin 
  dashed lines show $t_{\rm 3kT}/t_{\rm rc}=18 f_{\rm max}^{-1.3}$ and 
  $t_{\rm 3kT}/t_{\rm rc}=18 f_{\rm max}^{-1}$.  The
  dotted lines indicate the $t_{\rm 3kT}/t_{\rm rc} = 0.15$, which is
  suggested the minimum ratio between $t_{\rm cc}$ and $t_{\rm rc}$ by
  \citet{2004ApJ...604..632G}.
\label{fig:t_cc_w3_3kT}}
\end{figure}

We find that $t_{\rm cc}/t_{\rm rc} \propto f_{\rm max}^{-1}$, which
is inconsistent with a previous work by \citet{2004ApJ...604..632G},
who conclude that the core-collapse time is proportional to $f_{\rm
  max}^{-1.3}$.  This discrepancy can be attributed to their
Monte-Carlo code, which may not properly be able to follow the cluster
all the way to the moment of core collapse, but shows a core bounce at
an earlier instance. A hint of the early termination of their
calculations is visible in their Fig.\,2, where the Lagrangian radii
of the simulations with a mass function suddenly truncates. They
identify this moment as core collapse, but by inspection of our own
simulations the sudden break in the inner most Lagrangian radii is
generally associated with the formation of the first hard ($\sim
3kT$) binary.  We tested this hypothesis by analyzing the results of
our simulation up to the moment of the formation of the first hard
($3kT$) binary.  In figure \ref{fig:t_cc_w3_3kT}, we present the
formation time of the first $3kT$-binary, $t_{3kT}$, for models with
$W_0=3$ as a function of $f_{\rm max}$.  If the moment of core
collapse is identified by moment of formation of the first 1--$3kT$
hard binary, we find that $t_{3kT}/t_{\rm rc} \propto f_{\rm
  max}^{-1.3}$, which is consistent with the results of
\citet{2004ApJ...604..632G}.  In that case the core collapse time
saturates at $t_{\rm cc}/t_{\rm rc} \simeq 0.15$, which they suggested
to be associated with the minimum core-collapse time.

\section{Conclusions}

We performed a series of $N$-body simulations of star clusters with
various mass ranges and power laws of the mass function, and found
that the core-collapse time follows $t_{\rm cc}/t_{\rm rc} \propto
(m_{\rm max}/\langle m \rangle)^{-1}$ for clusters with $M/m_{\rm
  max}\apgt 100$.  When $M/m_{\rm max}\aplt 100$, this relation breaks
and $t_{\rm cc}$ saturates at $\sim 10 t_{\rm cross}$.  We
subsequently argue that star clusters with a mass function reach core
collapse on the dynamical-friction timescale of the most massive stars
(consistent with the results of \citet{2002ApJ...576..899P}).  We also
showed that the dynamical evolution of star clusters with a mass
function are driven by the relaxation timescale of the most massive
stars.  We define an effective number of stars $N_{\rm eff} = M/m_{\rm
  max}$ for which a multi-mass cluster shows a similar core collapse
behavior as in the equal-mass case.

When the mass of the most massive stars is relatively small ($M/m_{\rm
  max}\apgt 10^3$), we notice a pronounced peak in the core density
during the evolution.  As we increase the mass of the most massive
stars, $m_{\rm max}$, the peak density at the core bounce becomes
lower. We found that the binding energy of the hard binaries are a
good indicator for detecting the moment of the core collapse, even if
the density peak is ambiguous. We adopted that the critical binding
energy of hard binaries, with which the binary emits sufficient energy
to bounce the core, as the moment of
the core bounce.  We conclude that the binding energy criterion, 
$E_{\rm cr}\simeq 10f_{\rm max}kT$, gives
a more robust indicator for the core-collapse time compared to
inspection of the core density and radius.

\section*{Acknowledgments}
We thank Jeroen B\'{e}dorf for discussions and adapting the Sapporo2
library to sixth order, Alex Rimoldi for careful reading of the 
manuscript, and anonymous referee for useful comments on the manuscript. 
This work was supported by Postdoctoral
Fellowship for Research Abroad of the Japan Society for the Promotion
of Science (JSPS) and by Netherlands Research Council NWO
(\#639.073.803 [VICI] and \#614.061.608 [AMUSE]).  Numerical
computations were carried out on Cray XT4 and XC30 CPU-clusters at Center
for Computational Astrophysics (CfCA) of National Astronomical
Observatory of Japan and the Little Green Machine at Leiden
Observatory (NWO grant \#612.071.305).

\bibliographystyle{mn}
\bibliography{reference}

\label{lastpage}

\end{document}